%% file: paper.tex
\documentclass[12pt]{article}
\usepackage{jheppub}

\pdfoutput=1

\usepackage{amsmath,bbm,array,amsfonts,graphicx,wrapfig,lscape,float,mathtools,multirow,longtable}
\usepackage[dvipsnames]{xcolor}
\usepackage{array}

\input{pref}

\definecolor{darkspringgreen}{rgb}{0.09, 0.45, 0.27}
\definecolor{forestgreen}{rgb}{0.13, 0.55, 0.13}

\usepackage{array}
\usepackage{physics}
\usepackage{subcaption}
\usepackage{tikz,tikz-3dplot}
\usepackage[colorlinks=true]{hyperref}
\newcolumntype{C}[1]{>{\centering\let\newline\\\arraybackslash\hspace{0pt}}m{#1}}

\definecolor{yellow2}{rgb}{0.98, 0.80, 0.20}

\title{Twin Theories, Polytope Mutations and Quivers for GTPs }
\author[a,b,c]{Sebasti\'an Franco,}

\author[d,e]{Rak-Kyeong Seong}

\affiliation[a]{Physics Department, The City College of the CUNY\\
	160 Convent Avenue, New York, NY 10031, USA}
\affiliation[b]{Physics Program and \textsuperscript{$c$}Initiative for the Theoretical Sciences\\
	The Graduate School and University Center, The City University of New York\\
	365 Fifth Avenue, New York NY 10016, USA}
\affiliation[d]{
Department of Mathematical Sciences, and 
\textsuperscript{$e$}Department of Physics,\\ 
Ulsan National Institute of Science and Technology,\\
50 UNIST-gil, Ulsan 44919, South Korea
}
	
\emailAdd{sfranco@ccny.cuny.edu}
\emailAdd{seong@unist.ac.kr}

\preprint{
\begin{flushright}
UNIST-MTH-23-RS-01 \\
\end{flushright}
}

\abstract{We propose a unified perspective on two sets of objects that usually arise in the study of bipartite field theories. Each of the sets consists of a polytope, or equivalently a toric Calabi-Yau, and a quiver theory. We refer to the two sets of objects as original and twin. In the simplest cases, the two sides of the correspondence are connected by the graph operation known as untwisting. The democratic treatment that we advocate raises new questions regarding the connections between these objects, some of which we explore. 

With this motivation in mind, we establish a correspondence between the mutations of the original polytope and the twin quiver. This leads us to propose that non-toric twin quivers are naturally associated to generalized toric polygons (GTPs) and we explore various aspects of this idea. Supporting evidence includes global symmetries, the ability of twin quivers to encode the generalized $s$-rule, and the connection between the mutations of polytopes and of configurations of webs of 5-branes suspended from 7-branes. We introduce three methods for constructing twin quivers for GTPs. We also investigate the connection between twin quivers obtained using different toric phases. Twin quivers provide a powerful new perspective on GTPs. The ideas presented in this paper may represent a step towards the generalization of brane tilings to GTPs.
}

\begin{document}

\maketitle

%=================================================================
\section{Introduction} 
%=================================================================

Brane tilings encode the $4d$ quiver gauge theories on D3-branes probing toric Calabi-Yau (CY) 3-fold singularities \cite{Hanany:2005ve,Franco:2005rj,Franco:2005sm}. They are not only physical brane configurations related to the D-branes at singularities by T-duality, but they also significantly simplify the map between gauge theory and geometry. This family of theories was later generalized to the much broader class of {\it bipartite field theories} (BFTs), which are defined by bipartite graphs on Riemann surfaces and share many of the combinatorial properties and connections to toric geometry of brane tilings \cite{Franco:2012mm,Franco:2012wv,Franco:2013pg,Cremonesi:2013aba,Franco:2013ana,Franco:2014nca}. 

Interestingly, the same geometries and quiver theories are relevant for superconformal field theories in a different dimension. M-theory on CY$_3$ singularities engineers $5d$ superconformal field theories \cite{Seiberg:1996bd, Morrison:1996xf, Douglas:1996xp, Intriligator:1997pq}. Moreover, the BPS spectrum of such theories can be described by so-called $5d$ BPS quivers which, for toric CY$_3$’s, are indeed the theories associated to the corresponding brane tilings \cite{Closset:2019juk}. 

The study of BFTs leads to various quiver theories and polytopes. These objects and their mutations are naturally divided into two sets, which we denote original and twin, connected by the combinatorial operation of {\it untwisting}. For brane tilings, untwisting is intimately related to mirror symmetry \cite{Feng:2005gw}. Motivated by applications, some of these objects have been thoroughly studied, while others are rarely discussed in the literature. This is also the case for their mutations and the connections between them. One of the main points of this paper is to emphasize that the original and twin sides are on a symmetric footing and hence all these objects should be treated democratically. This new perspective will lead us to consider connections between the two sides that were previously overlooked. 

On a seemingly independent line of developments, {\it generalized toric polygons} (GTPs) have been introduced to describe brane configurations engineering $5d$ SCFTs \cite{Benini:2009gi,vanBeest:2020kou,VanBeest:2020kxw}. Given that GTPs extend the usual notion of toric geometry, it is natural to ask whether some of the objects mentioned above can be generalized to GTPs. For example, it would be particularly interesting to find a generalization of brane tilings to GTPs, since they could shed light on the BPS spectrum of the associated $5d$ theories. Remarkably, guided by the unified perspective advocated in the first part of the paper, we will propose that non-toric twin quivers are connected to GTPs. We will investigate various aspects of the proposed correspondence and show that these quivers are indeed useful tools, e.g. when analyzing the generalized $s$-rule. 

This paper is organized as follows. In Section \sref{section_BFTs_and_polytopes}, we review BFTs and several of the tools involved in their study, including perfect matchings, zig-zag paths and their connection to toric geometry via their moduli spaces. We also discuss quiver and polytope mutations. We finally present untwisting, which plays a central role throughout the paper. In Section \sref{section_democratic_perspective}, we consider an original BFT and the corresponding polytope, quiver mutation and polytope mutation. We also discuss how untwisting produces a twin BFT, which also comes with its associated polytope, quiver mutation and polytope mutation. We advocate that the original and twin sides should be regarded on an equal footing and therefore it is interesting to investigate the correspondence between objects and mutations on both sides. This democratic perspective underlies the rest of the paper. Section \sref{section_two_projections} explains how the toric diagrams for the moduli spaces of the original and twin BFTs are different projections of a single underlying polytope. Section \sref{section_GTPs} provides a brief summary of GTPs. In Section \sref{section_quiver_and_polytope_mutations}, we argue that quiver and polytope mutations are not independent operations. In fact, the mutation of the original polytope corresponds to the mutation of the twin quiver and vice versa. Section \sref{section_twin_quivers_for_GTPs} presents one of the main proposals of the paper: non-toric twin quivers are naturally associated to GTPs. This section also introduces two algorithms for determining the twin quiver for a given GTP.  Section \sref{section_direction_construction_quiver_for_GTP} presents a third method to directly compute the twin quiver from a GTP and discusses its limitations. Section \sref{section_twin_quivers_for_different_toric_phases} studies the relation between the twin quivers for a GTP constructed using different toric phases of the original theory. Section \sref{section_s-rule} explains how twin quivers capture the generalized $s$-rule. Section \sref{section_conclusions} summarizes our conclusions and outlines interesting directions for future research. Appendix \sref{appendix_details_models} presents additional details for some of the examples considered in the paper.

%=================================================================
\section{Graphs, gauge theories and polytopes}
%=================================================================

\label{section_BFTs_and_polytopes}

A significant fraction of the discussion in this and the coming two sections follows previous works (see \cite{Franco:2012mm} and references therein) albeit with a fresh perspective. In order to avoid repetition, we will keep the presentation of these subjects brief. We refer the interested reader to the original references for thorough discussions.

%=================================================================
\subsection{Bipartite field theories}
%=================================================================

A bipartite field theory is a $4d$ $\mathcal{N} = 1$ supersymmetric gauge theory defined by a bipartite graph $G$ embedded in a Riemann surface $\Sigma$, possibly including boundaries \cite{Franco:2012mm}.\footnote{A closely related class of theories was considered in \cite{Xie:2012mr}.}

A bipartite graph is a graph in which nodes can be colored white or black, such that nodes are only connected to nodes of the opposite color. Nodes can be further distinguished into internal and external. External nodes are those on the boundaries of the Riemann surface. We refer to the number of edges connected to a given node as its valence. In our construction, we will restrict to external nodes with valence one.

Faces are regions on the Riemann surface that are surrounded by edges and/or boundaries. They can also be classified into external or internal, with external faces being those whose perimeter includes at least one boundary.

The dictionary connecting a bipartite graph on a Riemann surface and a BFT is summarized in Table \ref{table_1}.

%=================================================================
\begin{table}[H]
\begin{center}
\begin{tabular}{ | m{4cm} | m{9.5cm}|  }
\hline 
{\bf Graph} & {\bf BFT} \\ \hline \hline
Internal Face & $U(N)$ gauge symmetry group \\ \hline 
External Face & $U(N)$ global symmetry group \\ \hline
Edge between faces $i$ and $j$ & 
Chiral superfield in the bifundamental representation of groups $i$ and $j$  (adjoint representation if $i = j$). The chirality, i.e. orientation, of the bifundamental is such that it goes clockwise around white nodes and counter-clockwise around black nodes. \\ \hline
$k$-valent node &
Superpotential monomial consisting of $k$ chiral superfields. Its sign is $+$/$-$ for a white/black node, respectively. \\ \hline
\end{tabular}
\caption{The dictionary relating bipartite graphs on Riemann surfaces to BFTs.}
\label{table_1}
\end{center}
\end{table}
%=================================================================

\begin{figure}[ht!]
\begin{center}
\resizebox{0.75\hsize}{!}{
  \includegraphics[trim=0mm 0mm 0mm 0mm, width=8in]{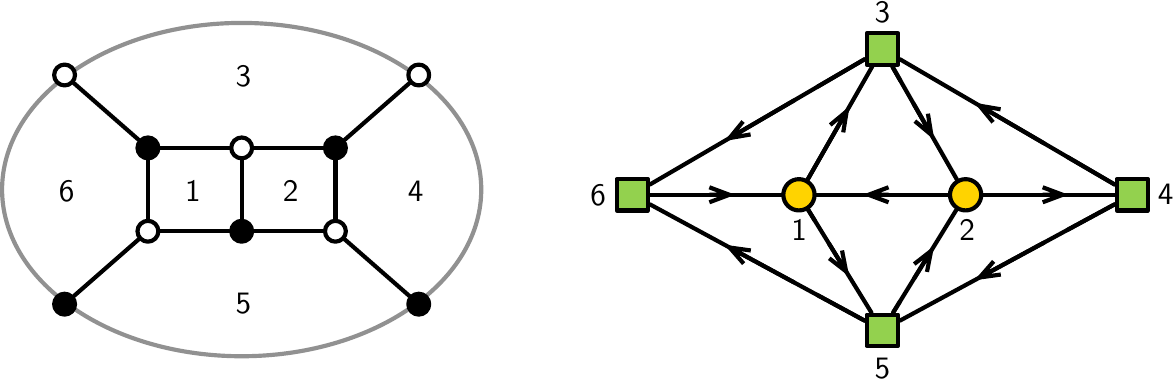}
}
\caption{A bipartite graph on a disk and its dual quiver. Every plaquette in this quiver corresponds to a node in the original graph and, hence, to a superpotential term.
\label{example_general_BFT}
}
\end{center}
\end{figure}

We can equivalently think about these theories in terms of a quiver dual to the
graph, as shown in \fref{example_general_BFT}. This quiver is such that its plaquettes, i.e. the minimal oriented closed loops, are in one-to-one correspondence with nodes in the graph and, therefore, with terms in the superpotential.

%=================================================================
%=================================================================

It is possible, and indeed physically well motivated, to consider the case in which the ranks of the $U(N_i)$ gauge or global symmetry groups associated to different faces are not equal. For simplicity, we will not consider this possibility in this article. We will often refer to BFTs, namely to theories described by bipartite graphs, as {\it toric phases}. The first part of the paper will be dedicated to toric phases. Starting in Section \sref{section_quiver_and_polytope_mutations}, we will also consider non-toric phases, which can be generated from toric phases by quiver mutation. For non-toric phases, not only the Lagrangian cannot be captured by bipartite graphs, but also different ranks are unavoidable. 

\bigskip

%=================================================================
\paragraph{Perfect matchings.} 
%=================================================================
A perfect matching $p$ is a subset of the edges in $G$ such that:\footnote{When $G$ has boundaries, the objects defined here are often called {\it almost perfect matchings}.} 
\begin{itemize}
\item Every internal node is the endpoint of exactly one edge in $p$.
\item Every external node belongs to either one or zero edges in $p$.
\end{itemize}

Perfect matchings connect bipartite graphs and the corresponding BFTs to toric geometry, as we explain in Section \sref{section_two_projections}. For doing so, it is convenient to consider the following map between chiral fields in the quiver $X_i$, i.e. edges in $G$, and perfect matchings $p_\mu$ 
\beq
X_i = \prod_{\mu=1}^c p_\mu^{P_{i\mu}} \, ,
\label{X_pm_map}
\eeq
where $c$ is the total number of perfect matchings, and $P_{i\mu}$ is equal to $1$ if the edge in the bipartite graph associated to the chiral field $X_i$ is contained in $p_\mu$ and zero otherwise \cite{Franco:2005rj,Franco:2012mm}, i.e.
\beq
P_{i\mu}=\left\{ \begin{array}{ccccc} 1 & \rm{ if } & X_i  & \in & p_\mu \\
0 & \rm{ if } & X_i  & \notin & p_\mu
\end{array}\right.
\label{Xi_to_pmu}
\eeq
Perfect matchings are particular useful because they are efficient ways to determine them. These method are base on the Kasteleyn matrix \cite{Franco:2005rj} and generalizations in the case of graphs with boundaries \cite{Franco:2012mm}.

\bigskip

\begin{figure}[H]
\begin{center}
\resizebox{0.4\hsize}{!}{
  \includegraphics[trim=0mm 0mm 0mm 0mm, width=8in]{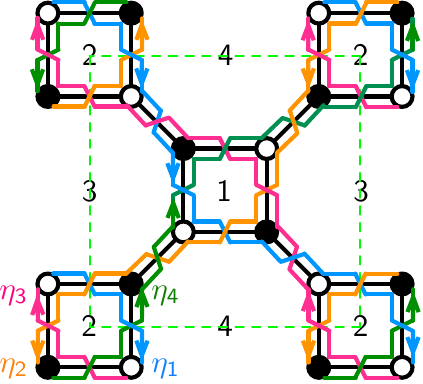}
}
\caption{Brane tiling for $F_0$ and its zig-zag paths. Dashed green lines indicate the unit cell.
\label{zig_zags_F0}
}
\end{center}
\end{figure}

%=================================================================
\paragraph{Zig-zag paths.} 
%=================================================================
Zig-zag paths are oriented paths in a bipartite graph embedded in a Riemann surface that alternate between turning maximally right and maximally left at white and black nodes, respectively. They can be nicely implemented in terms of a double line notation for edges \cite{Franco:2012mm}, such that two zig-zag paths go over every edge in opposite directions, crossing at the middle point. \fref{zig_zags_F0} shows a simple bipartite graph corresponding to the $F_0$ geometry and its zig-zag paths.

%=================================================================
%=================================================================

Zig-zag paths play a central role in the study of bipartite graphs for several reasons. First, it is possible to reconstruct the graph $G$ from knowledge of its zig-zags \cite{Hanany:2005ss}. Second, the conditions for consistency/irreducibility of a bipartite graph can be elegantly formulated in terms of properties of its zig-zag paths \cite{MR2908565}. Finally, similarly to perfect matchings, zig-zag paths provide a powerful tool for connecting BFTs to their moduli spaces.

\bigskip

%=================================================================
\paragraph{BFTs and toric geometry.} 
%=================================================================

BFTs are intimately related to two non-compact toric CYs, which correspond to their moduli and master spaces \cite{Franco:2012mm}. They are therefore associated to two convex lattice polytopes, which are the toric diagrams of the two CYs. The dimensionality of these CYs, equivalently the dimensionality of their toric diagrams, can be directly determined from general properties of the bipartite graph and the Riemann surface it is embedded in. We refer the interested reader to \cite{Franco:2013nwa} for an extensive discussion of this issue.

%=================================================================
\subsection{The special case of $\mathbb{T}^2$: $2d$ toric diagrams, $(p,q)$-webs and $5d$ theories}
%=================================================================

\label{section_special_case_T2}

At various points of the paper we will focus on brane tilings, namely BFTs defined by bipartite graphs on $\mathbb{T}^2$, so let us discuss them in further detail. The moduli space for such a theory is a toric CY$_3$ $X_\Delta$, where $\Delta$ denotes the corresponding $2d$ toric diagram. 

\begin{figure}[H]
\begin{center}
\resizebox{0.55\hsize}{!}{
  \includegraphics[trim=0mm 0mm 0mm 0mm, width=8in]{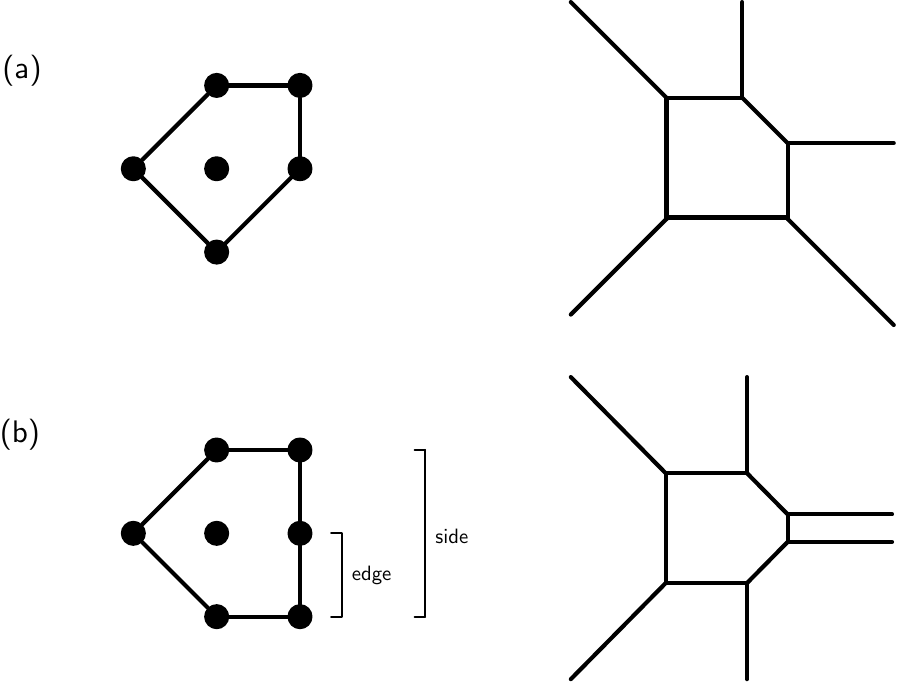}
}
\caption{Two toric diagrams and their dual $(p,q$)-webs.
\label{toric_diagrams_and_webs}
}
\end{center}
\end{figure}

$2d$ toric diagrams are closely related to $(p,q)$-webs (see \cite{Aharony:1997ju,Aharony:1997bh,Leung:1997tw} for detailed presentations). Given a triangulation of a toric diagram, we obtain a $(p,q)$-web by graph dualization, as illustrated in \fref{toric_diagrams_and_webs}. Every line in the web corresponds to a $(p,q)$ 5-brane. The information regarding lengths of lines in the $(p,q)$-web is lost when going to the corresponding toric diagram.

For discussing the boundary of a toric diagram, it will be useful to introduce the notions of {\it sides} and {\it edges}. As usual, we refer to a side as the line connecting two consecutive corners of the toric diagram. Within a given side, an edge is a segment between two consecutive points in the toric diagram. The distinction between the two becomes relevant in the case of sides consisting of more than one edge. \fref{toric_diagrams_and_webs}.a shows a toric diagram in which every side involves a single edge. On the other hand, \fref{toric_diagrams_and_webs}.b shows a toric diagram in which one of the sides consists of two edges. The corresponding $(p,q)$-web therefore has two parallel external legs.

%=================================================================
%=================================================================

The orientation of every external leg $\tilde{i}$ of the $(p,q)$-web, which represents the outward normal vector to the corresponding boundary edge $\tilde{i}$ of the toric diagram, takes the form
\beq
\eta_{\tilde{i}}=(p_{\tilde{i}},q_{\tilde{i}}) \in \mathbb{Z}^2
\, .
\eeq

Every edge of the toric diagram or, equivalently, every leg of the $(p,q)$-web is in one-to-one correspondence with a zig-zag path of the associated brane tiling. The vectors $\eta_{\tilde{i}}$ are the winding numbers of the zig-zags along the two fundamental cycles of $\mathbb{T}^2$ \cite{Feng:2005gw}. Clearly, sides with multiple edges correspond to parallel legs in the dual $(p,q)$-web. This also means that the associated brane tiling contains more than one zig-zag path with the same winding numbers.

Many of the objects previously discussed beautifully interplay in the context of $5d$ theories. The low energy limit of M-theory on $\mathbb{R}^{1,4}$ times the local CY$_3$ associated to a toric diagram gives rise to a $5d$ $\mathcal{N}=1$ theory \cite{Morrison:1996xf}. The same theory is also engineered by the corresponding $(p,q)$-web \cite{Aharony:1997ju,Aharony:1997bh}. The BPS spectrum of such a $5d$ theory is encoded in a quiver theory that is precisely the one defined by the corresponding brane tiling \cite{Closset:2019juk}.

%=================================================================
\subsection{Quiver mutation} 
%=================================================================

This section reviews the mutation of quiver theories, equivalently Seiberg duality \cite{Seiberg:1994pq}.\footnote{Throughout this paper, we will use the term {\it quiver theory} to refer to the full gauge theory, namely not only to its quiver but also its superpotential.} Let us consider acting with the mutation on a node $j$ of the quiver, which we assume does not have adjoint chiral fields. Below we summarize how the quiver and the superpotential transform.

%=================================================================
\subsection*{Quiver} 
\begin{figure}[H]
\begin{center}
\resizebox{0.7\hsize}{!}{
  \includegraphics[trim=0mm 0mm 0mm 0mm, width=8in]{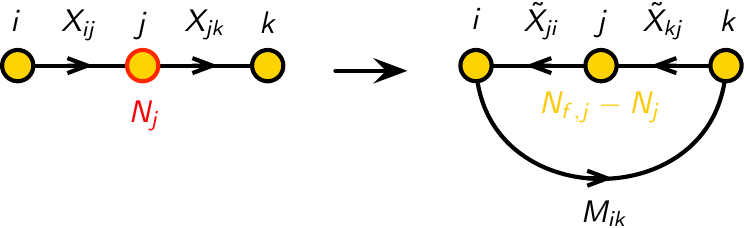}
}
\caption{Quiver mutation.
\label{quiver_mutation}
}
\end{center}
\end{figure}

%=================================================================
\begin{itemize}
\item \textbf{Flavors.} Chiral fields connected to the mutated node transform as follows. Incoming fields $X_{ij}$ and outgoing fields $X_{jk}$ are replaced by dual flavors pointing in the opposite directions $\tilde{X}_{ji}$ and $\tilde{X}_{kj}$, respectively.

\item \textbf{Mesons.} 
Mesons $M_{ik}$ are added for every $2$-path consisting of flavors $X_{ij}$ and $X_{jk}$.

\item \textbf{Ranks.} 
The rank of node $j$ transform as $N_j^\prime = N_{f,j} - N_j$,
where $N_{f,j}$ is the number of flavors at the node $j$, given by 
\beal{es10a1}
N_{f,j} = \sum_{i} n_{ij} N_i = \sum_{k} n_{jk} N_k \, ,
\eea
where $n_{ij}$ is the number of bifundamental fields between nodes $i$ and $j$.
\end{itemize}
The above transformation on the quiver is illustrated in \fref{quiver_mutation}. 

%=================================================================
%=================================================================

%=================================================================
\subsection*{Superpotential}
%=================================================================

\begin{itemize}
\item Instances of $X_{ij} X_{jk}$ in the original superpotential are replaced by the meson $M_{ik}$.
\item A cubic term of the form $M_{ik} \tilde{X}_{kj} \tilde{X}_{ji}$ is added for every meson.
\end{itemize}
The transformation of the superpotential can result in mass terms, i.e. quadratic terms. Massive fields can be integrated out using the equations of motion. Due to gauge invariance, to admit a mass term chiral fields must be adjoints or pairs of fields forming a bidirectional arrow. However, not every adjoint or bidirectional arrow necessarily participates in a mass term. Such fields should not be removed from the quiver.

%=================================================================
\subsubsection{Quiver mutations connecting toric phases} 
%=================================================================

\begin{figure}[H]
\begin{center}
\resizebox{0.6\hsize}{!}{
  \includegraphics[trim=0mm 0mm 0mm 0mm, width=8in]{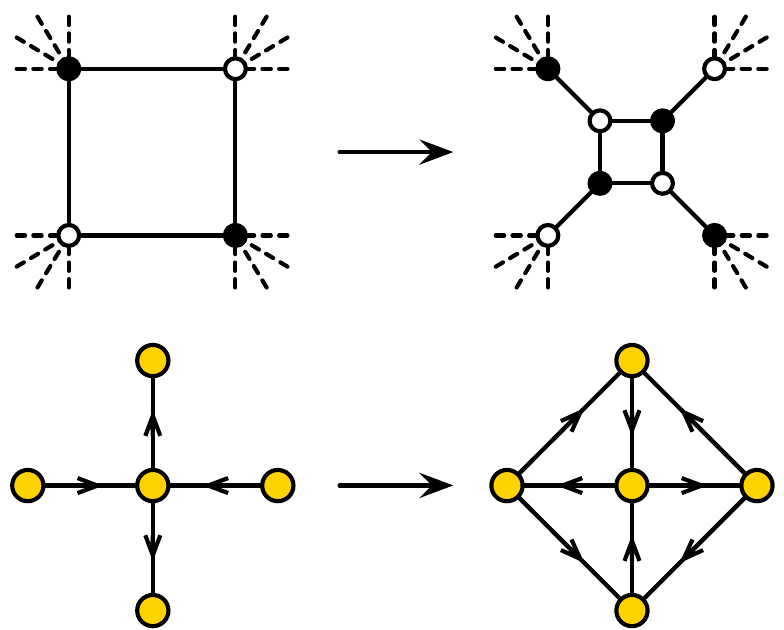}
}
\caption{A quiver mutation connecting two toric phases maps to a square move in the bipartite graph.
\label{square_move}
}
\end{center}
\end{figure}

It is interesting to discuss in further detail the case of quiver mutations in which the two dual theories are BFTs on the same Riemann surface. This constraint implies that we can only dualize $N_f = 2 N_c$  gauge groups, where $N_c = N$, the common rank of all gauge and global symmetry groups. This class of gauge groups are represented by internal square faces in the bipartite graph. The quiver mutation of such gauge groups is beautifully realized by a local transformation of the graph denoted {\it square move}, which is shown in \fref{square_move} \cite{Franco:2005rj,Franco:2012mm}. In this case, the rank of the gauge group remains constant.

%=================================================================
%=================================================================

%=================================================================
\subsection{Polytope Mutation} 
%=================================================================

\label{section_polytope_mutation}

This section is devoted to {\it polytope mutation}. This operation can be defined for lattice polytopes in arbitrary dimensions (see e.g. \cite{Higashitani:2019vzu}). It is reasonable to expect that such general definition will play a role when the ideas presented later in this paper are considered in full generality. We leave this for future work. Motivated by specific questions that will be addressed in Sections \ref{section_twin_quivers_for_GTPs} to \ref{section_s-rule}, we will now focus on the case of 2-dimensional polytopes, which can be interpreted as toric diagrams of CY 3-folds. 

We can discuss toric diagrams in terms the outward pointing normal vectors. We will now consider one such vector for each side of the toric diagram, regardless of how many edges it contains. Following the notation of Section \sref{section_special_case_T2}, we define the intersection number between two sides of the toric diagram with normal vectors $\eta_{\tilde{i}}$ and $\eta_{\tilde{j}}$ as follows
\beq
\langle \eta_{\tilde{i}},\eta_{\tilde{j}} \rangle = \det \left(\begin{array}{cc} p_{\tilde{i}} & q_{\tilde{i}} \\ p_{\tilde{j}} & q_{\tilde{j}} \end{array}\right)
\, .
\label{intersection_sides}
\eeq

The normal vectors need to satisfy the following relation
\beq
\sum_{\tilde{i}} N_{\tilde{i}} \eta_{\tilde{i}} =0 \, ,
\label{sum_charges}
\eeq
where $N_{\tilde{i}}$ is the number of edges in a given side, i.e. the multiplicity of legs in the $(p,q)$-web that have the same $(p_{\tilde{i}},q_{\tilde{i}})$-charges. Equation \eref{sum_charges} is simply the equilibrium condition for the $(p,q)$-web.

Mutating a polytope with respect to a side $\tilde{j}$, corresponds to changing the normal vectors as follows
\beq
\begin{array}{cll}
& \eta_{\tilde{j}}'  =  -\eta_{\tilde{j}} &  \\[.2cm]
\tilde{i} \neq \tilde{j}: \ \  & \eta_{\tilde{i}}' = \eta_{\tilde{i}} + \langle \eta_{\tilde{j}},\eta_{\tilde{i}} \rangle \eta_j \ \ \ \ & \mbox{for } \langle \eta_{\tilde{j}},\eta_{\tilde{i}} \rangle > 0 \\[.2cm]
& \eta_{\tilde{i}}' = \eta_{\tilde{i}} & \mbox{otherwise}
\end{array}
\label{mutation_polytope_1}
\eeq
In order to satisfy \eref{sum_charges} after the mutation, $N_{\tilde{j}}$ must transform according to
\beq
N_{\tilde{j}}' = \sum_{\tilde{i} \in E_+} N_{\tilde{i}} \langle \eta_{\tilde{j}},\eta_{\tilde{i}} \rangle - N_{\tilde{j}}
\, ,
\label{mutation_polytope_2}
\eeq
where $E_{+}$ is the set of sides of the toric diagram with $\langle \eta_{\tilde{j}},\eta_{\tilde{i}} \rangle > 0 $. We will later argue that it is more appropriate to regard configurations in which some $N_{\tilde{i}}>1$ as GTPs rather than as ordinary toric diagrams.

\begin{figure}[ht!!]
\begin{center}
\resizebox{0.8\hsize}{!}{
\includegraphics[height=6cm]{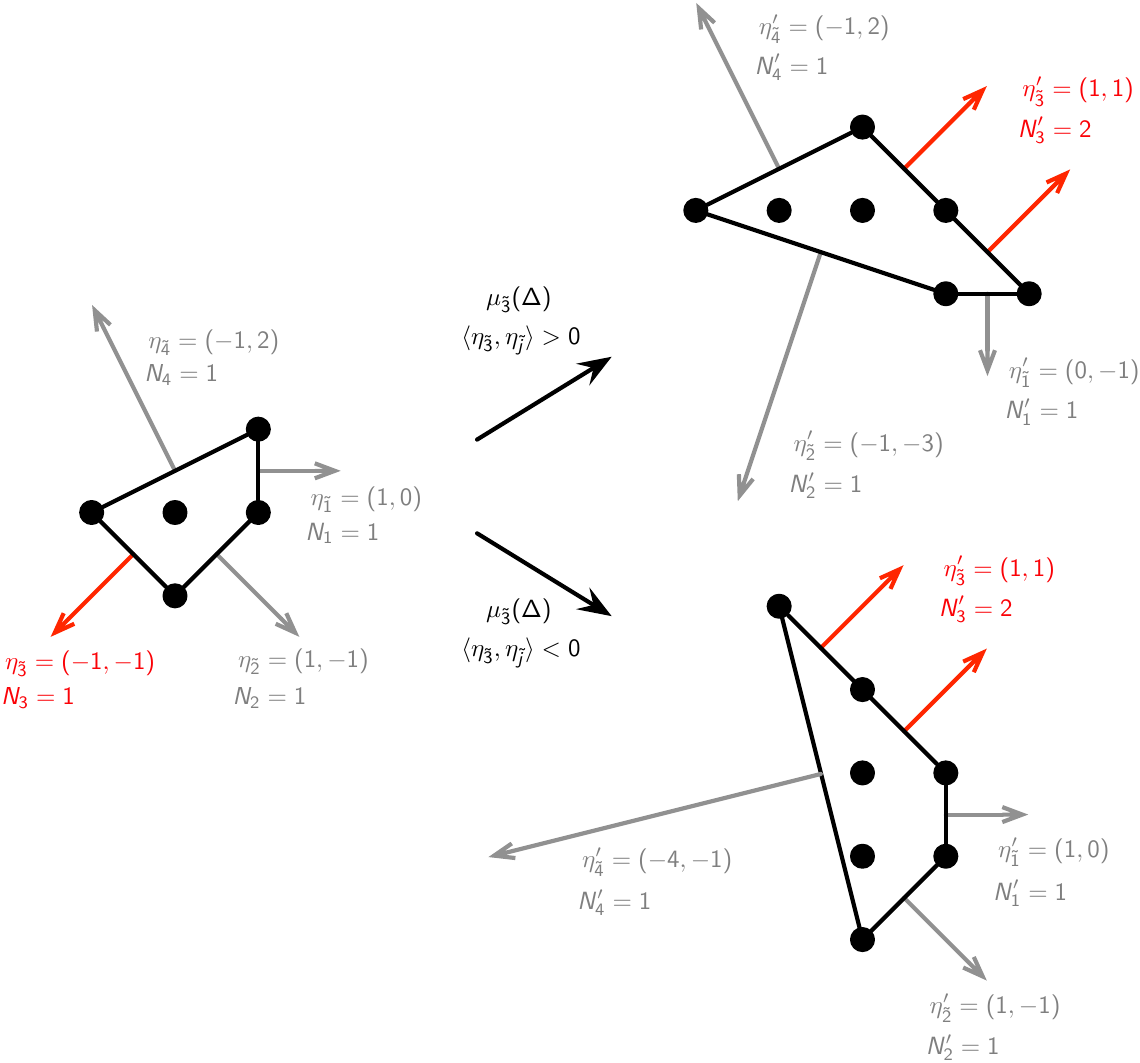} 
}
\caption{
The toric diagram for $dP_1$ before and after the polytope mutation on $\eta_{\tilde{1}}$, where we show both the mutation with the condition $\langle \eta_{\tilde{1}},\eta_{\tilde{i}} \rangle > 0 $ and $\langle \eta_{\tilde{1}},\eta_{\tilde{i}} \rangle < 0 $.
\label{f_polytopemutation}}
\end{center}
\end{figure}

There is an equivalent transformation in which, for a given mutated side $\tilde{j}$ with charge $\eta_{\tilde{j}}$, the roles of the $\eta_{\tilde{i}}$'s with $\langle \eta_{\tilde{j}},\eta_{\tilde{i}} \rangle >0$ and $\langle \eta_{\tilde{j}},\eta_{\tilde{i}} \rangle < 0$ are exchanged in \eref{mutation_polytope_1}. More precisely, this corresponds to replacing $\langle \eta_{\tilde{j}},\eta_{\tilde{i}} \rangle$ by $-\langle \eta_{\tilde{j}},\eta_{\tilde{i}} \rangle$ everywhere in \eqref{mutation_polytope_1}. \fref{f_polytopemutation} illustrates a polytope mutation on the toric diagram of $dP_1$, which acts on a side $\tilde{j}$ such that its rank does not change, i.e. $N_{\tilde{j}}=N_{\tilde{i}}'$. We show the results of the mutations that modify either the sides with $\langle \eta_{\tilde{j}},\eta_{\tilde{i}} \rangle > 0$ or $\langle \eta_{\tilde{j}},\eta_{\tilde{i}} \rangle > 0$. Both results are equivalent up to $SL(2,\mathbb{Z})$ transformations. From now on, we denote the toric diagram before the polytope mutation as $\Delta$ and after the polytope mutation on $\eta_{\tilde{j}}$ as $\mu_{\tilde{j}} (\Delta)$.

%=================================================================
\begin{figure}[H]
\begin{center}
\resizebox{0.8\hsize}{!}{
  \includegraphics[trim=0mm 0mm 0mm 0mm, width=8in]{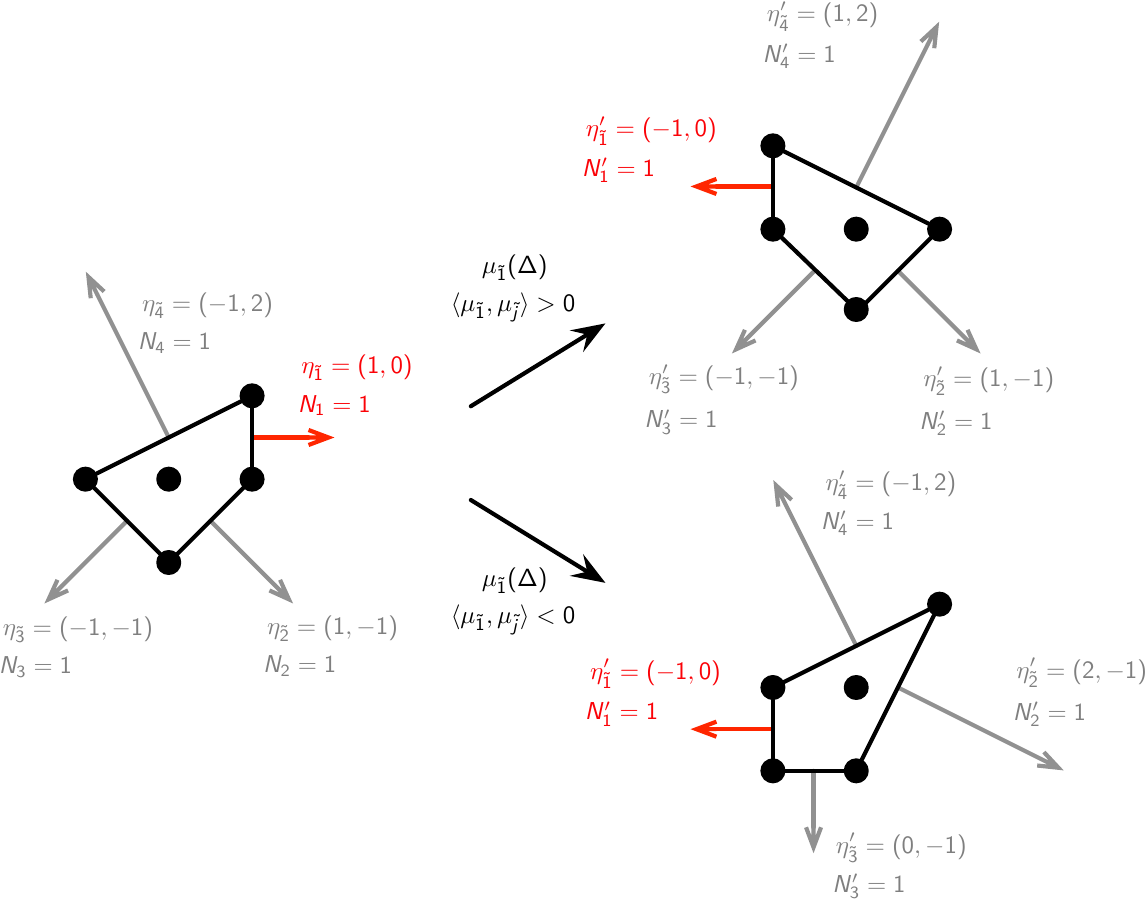}
}
\caption{Polytope mutation on the toric diagram of $dP_1$. The mutated side is such that its rank does not change.
\label{f06}
}
\end{center}
\end{figure}
%=================================================================

\fref{f_polytopemutation} shows the mutation of the same polytope acting on a different side. This time, the number of edges in the mutated side goes from one to two.

%=================================================================
%=================================================================

%=================================================================
\subsection{Untwisting} 
%=================================================================

\begin{figure}[H]
\begin{center}
\resizebox{0.75\hsize}{!}{
  \includegraphics[trim=0mm 0mm 0mm 0mm, width=8in]{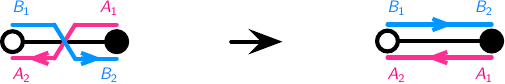}
}
\caption{The untwisting map.
\label{figure_untwisting}
}
\end{center}
\end{figure}

A central ingredient of our discussion will be the operation denoted {\it untwisting} which is illustrated in \fref{figure_untwisting} (see e.g \cite{Feng:2005gw,Franco:2012mm} for further discussion).

%=================================================================
%=================================================================

This operation exchanges zig-zag paths and faces. When doing so, it preserves the bipartite graph but it generically changes the Riemann surface in which it is embedded. The graph $G$ on the new Riemann surface $\tilde{\Sigma}$ can be interpreted as defining a new BFT, to which we will refer as $\tilde{Q}$. We can summarize the action of the untwisting map as follows:
\beal{es30a1}
\begin{tabular}{ccc}
$Q$ on $\Sigma$ & $\leftrightarrow$ & $\tilde{Q}$ on $\tilde{\Sigma}$ \\[.15cm]
zig-zag path $\eta_{\tilde{i}}$ & $\leftrightarrow$ & face/gauge group $\tilde{i}$\\[.15cm]
face/gauge group $i$ & $\leftrightarrow$ & zig-zag path $\eta_{i}$ \\
\end{tabular}
\, .
\eea

%=================================================================
\section{A democratic tale of polytopes, quivers and mutations} 
%=================================================================

\label{section_democratic_perspective}

In the previous section we reviewed BFTs and their associated polytopes. We also introduced mutations for both of them. These objects and transformations can be naturally organized into pairs, where untwisting serves as the bridge taking back and forth between them.

In order to organize our discussion, we will refer to the two sets of objects and mutations connected by untwisting as the {\it original} and {\it twin} sides. Since there is a symmetry that exchanges both sides, the distinction between what we call original or twin is arbitrary.

We will focus on the following objects:

\bigskip

\noindent{\underline{{\bf Original}}}

\begin{itemize}
\item $Q$: original BFT, i.e. both quiver and superpotential, defined by $G$ on $\Sigma$. Later we will extend the class of objects we consider, allowing also for $Q$ to be a non-toric phase of these theories.
\item $\Delta$: toric diagram of the moduli space of the theory $Q$.
\item $\mu_i (Q)$: theory obtained by acting with a quiver mutation on node $i$ of $Q$. 
\item $\mu_{\tilde{i}} (\Delta)$: polytope obtained by acting with a polytope mutation on side $\tilde{i}$ of $\Delta$.\footnote{For simplicity, we use $\mu$ to indicate both quiver and polytope mutations. The meaning should always be clear from the object they act on.} 
\end{itemize}

\bigskip

The analogous objects on the twin side are:

\bigskip

\noindent{\underline{{\bf Twin}}}

\begin{itemize}
\item $\tilde{Q}$: BFT obtained from $Q$ by untwisting, namely defined by $G$ on $\tilde{\Sigma}$. As previously mentioned, we will later generalize the discussion to include non-toric phases.
\item $\tilde{\Delta}$: toric diagram of the moduli space of the theory $\tilde{Q}$.
\item $\mu_{\tilde{i}} (\tilde{Q})$: theory obtained by acting with a quiver mutation on node $\tilde{i}$ of $\tilde{Q}$. Here we have used the fact that untwisting exchanges zig-zags and faces, so the nodes of $\tilde{Q}$ correspond to zig-zags of the original theory.
\item $\mu_i (\tilde{\Delta})$: polytope obtained by acting with a polytope mutation on side $i$ of $\tilde{\Delta}$. Here we have used the fact the zig-zags of $\tilde{Q}$, i.e. the edges of $\tilde{\Delta}$, correspond to the nodes/faces $i$ of $Q$.
\end{itemize}

\bigskip

\begin{figure}[H]
\begin{center}
\resizebox{0.75\hsize}{!}{
  \includegraphics[trim=0mm 0mm 0mm 0mm, width=8in]{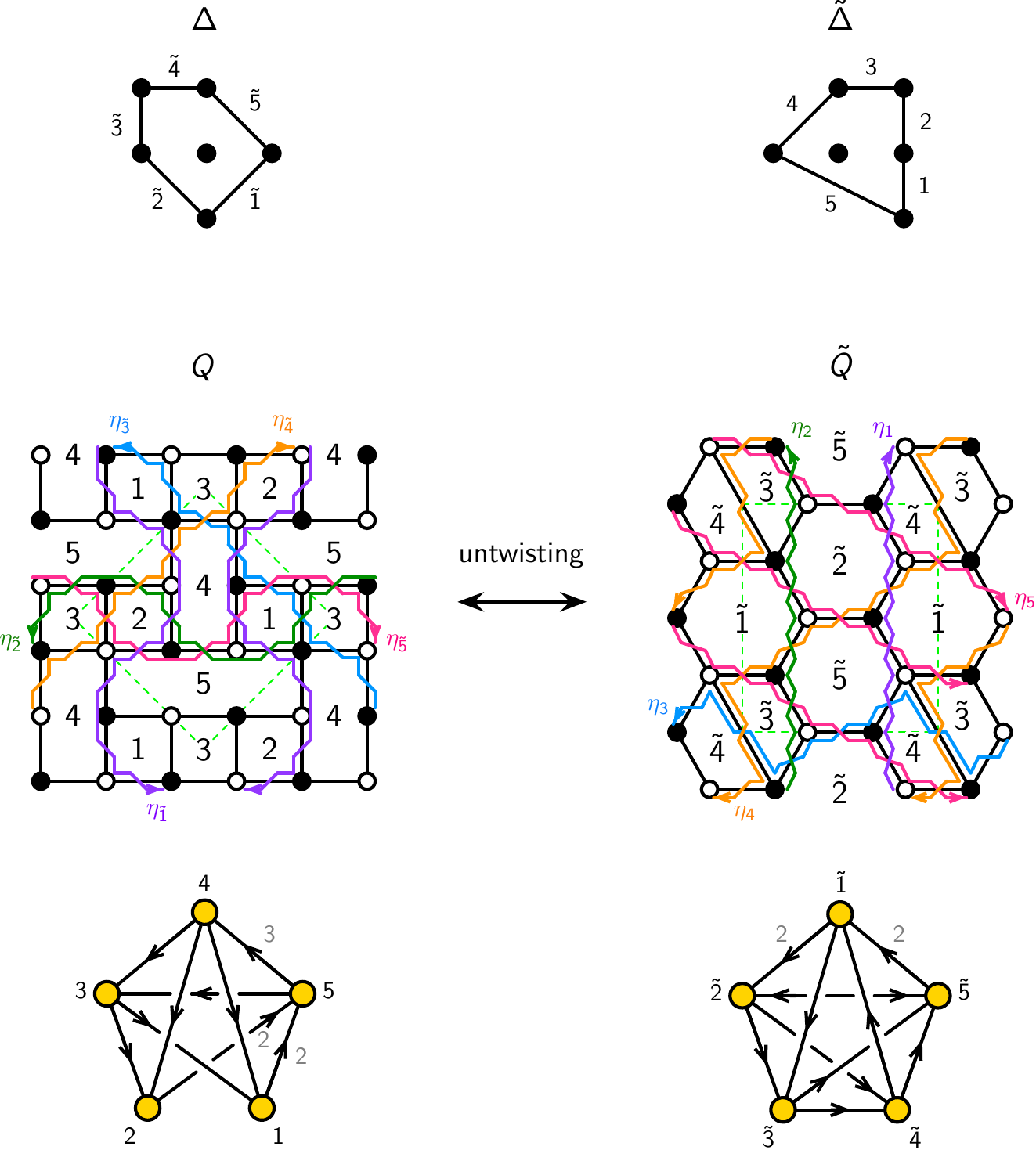}
}
\caption{An example of original and twin polytopes and quiver theories. The two sides of the correspondence are connected by untwisting.
\label{two_sides_of_the_correspondence}
}
\end{center}
\end{figure}

\fref{two_sides_of_the_correspondence} illustrates these ideas with an explicit example. To be able to properly represent the two sides, we have chosen an example in which the original and twin BFTs are given by brane tilings on $\mathbb{T}^2$ and, as a result, both polytopes are 2-dimensional. In this case, this is guaranteed because the polytopes are reflexive, i.e. they have a single internal point. Generically, however, even if one of the two sides corresponds to a brane tiling and a 2-dimensional polytope, the Riemann surface $\tilde{\Sigma}$ for the BFT obtained by untwisting might have a different genus and the corresponding toric diagram might not be 2-dimensional. The original and twin theories correspond to $dP_2$ and $PdP_2$, respectively.

%=================================================================
%=================================================================

The two models in \fref{two_sides_of_the_correspondence} are such that the number of faces and zig-zag paths in the BFTs, equivalently the area and perimeter of the corresponding polytopes, are equal. This property is not generic and it also follows from the fact that these polytopes are reflexive.

Many of the objects and transformations discussed above have been studied at length in the literature, within specific contexts. For example, $Q$ and $\Delta$ have been considered in the context of gauge theories on D-branes probing toric singularities \cite{Morrison:1998cs,Beasley:1999uz,Feng:2000mi,Feng:2001xr,Beasley:2001zp,Feng:2001bn,Franco:2005rj} or, more generally, BFTs and their associated geometries \cite{Franco:2012mm,Franco:2012wv,Franco:2014nca}. In addition, the invariance of $\Delta$ under $\mu_i (Q)$ has been thoroughly investigated, since it corresponds to the invariance of the moduli space of $Q$ under Seiberg duality \cite{Feng:2000mi,Feng:2001xr,Beasley:2001zp,Feng:2001bn,Franco:2005rj}. Equivalently, it underlies the fact that multiple gauge theories correspond to the same underlying geometry. Finally, $\tilde{Q}$, or more precisely the embedding of $G$ into a new Riemann surface obtained by untwisting, plays a central role in mirror symmetry for D-branes at singularities \cite{Feng:2005gw}. The associated BFTs were studied further in \cite{Franco:2012mm,Franco:2012wv,Franco:2014nca} and in relation with specular duality \cite{Hanany:2012vc}.

One of the main points that we would like to advocate in this paper is that the original and twin sides, as well as the connections between them, {\it must be studied in a unified way}.  More broadly, this story falls into the general spirit underlying several recent developments in physics: combinatorial objects, associated positive/convex geometries, the existence of more than one perspective (with each of them making different aspects manifest or simple), the existence of certain objects interpolating between them, etc.

The democratic perspective that we propose leads to new questions that are not often considered. For example, it is natural to ask what a mutation of the original quiver theory $\mu_i(Q)$ maps to on the twin side. This problem will be addressed in Section \sref{section_quiver_and_polytope_mutations}.

%=================================================================
\section{Two projections of the matching polytope} 
%=================================================================

\label{section_two_projections}

We now discuss the connection between $\Delta$ and $\tilde{\Delta}$ and how they are obtained as two different projections of a single underlying polytope. Throughout this section, we will assume that $Q$ and $\tilde{Q}$ are toric phases, namely that they are BFTs defined by bipartite graphs. In turn, as previously mentioned, this implies that $\Delta$ and $\tilde{\Delta}$ can be interpreted as toric diagrams. In Section \sref{section_twin_quivers_for_GTPs}, we will revisit this perspective and explain under which circumstances we consider that it is more appropriate to regard these polytopes, more generally, as GTPs. 

As explained in section \sref{section_democratic_perspective}, $\Delta$ and $\tilde{\Delta}$ are the toric diagrams for the moduli spaces of the BFTs $Q$ and $\tilde{Q}$, respectively.\footnote{Here we are interested in the mesonic moduli spaces.} Moduli spaces can be constructed in two stages. The first step is the determination of the {\it master space}, which is defined as the space of solutions to vanishing F-term equations \cite{Forcella:2008eh,Forcella:2008bb}. The moduli spaces of BFTs are toric and perfect matchings are in one-to-one correspondence with fields in their gauged linear sigma model (GLSM) description. Indeed, if chiral fields are expressed in terms of perfect matchings as in \eqref{Xi_to_pmu}
\beq
X_i = \prod_{\mu=1}^c p_\mu^{P_{i \mu}} \, ,
\label{map_pm_to_X}
\eeq
then F-term equations are automatically satisfied (see e.g. \cite{Franco:2012mm} for an explanation). The master space of a BFT is therefore naturally parameterized in terms of perfect matchings. In GLSM language, F-term conditions can be translated into certain $U(1)$ charges of the perfect matchings, which are encoded in a charge matrix $Q_F$ defined as
\beq
Q_F =Ker \, P \, .
\label{Q_F}
\eeq
The toric diagram of the master space, which we will denote $\Delta_M$, is given by $Ker \, Q_F$, which is the $P$-matrix. In the mathematics literature, $\Delta_M$ is also referred to as the {\it matching polytope} \cite{MR2525057}. Since both theories are defined by the same graph and differ only in its embedding into a Riemann surface, they share the same master space and hence $\Delta_M$.\footnote{Theories related in this way have been denoted {\it specular duals} in \cite{Hanany:2012vc}.}

Let us illustrate our discussion with the examples in \fref{two_sides_of_the_correspondence}. The toric diagram for the master space $\Delta_M$ is common to both theories and is given by the $P$-matrix. To visualize the resulting geometry, it is convenient to row-reduce $P$, which in this case becomes
\beq
G_{mast}=
\left(
\begin{array}{ccccc|cccccc}
\ p_1 \ & \ p_2 \ & \ p_3 \ & \ p_4 \ & \ p_5 \ & \ q_1 \ & \ q_2 \ & \ q_3 \ & \ q_4 \ & \ q_5 \ & \ q_6 \
\\
\hline
 0 & 1 & 1 & 0 & 0 & 1 & 0 & 0 & 0 & 0 & 1 \\
 1 & 0 & 0 & 0 & 0 & 1 & 0 & 0 & 0 & 1 & 0 \\
 1 & 1 & 0 & 0 & 0 & 0 & 0 & 0 & 1 & 0 & 0 \\
 0 & 0 & 1 & 0 & 0 & 0 & 0 & 1 & 0 & 0 & 0 \\
 0 & 0 & -1 & 0 & 0 & -1 & 1 & 0 & 0 & 0 & 0 \\
 0 & -1 & -1 & 0 & 1 & 0 & 0 & 0 & 0 & 0 & 0 \\
 -1 & 0 & 1 & 1 & 0 & 0 & 0 & 0 & 0 & 0 & 0 \\
\end{array}
\right) \, .
\label{G_mast}
\eeq

\begin{figure}[ht!!]
\begin{center}
\resizebox{0.55\hsize}{!}{
  \includegraphics[trim=0mm 0mm 0mm 0mm, width=9in]{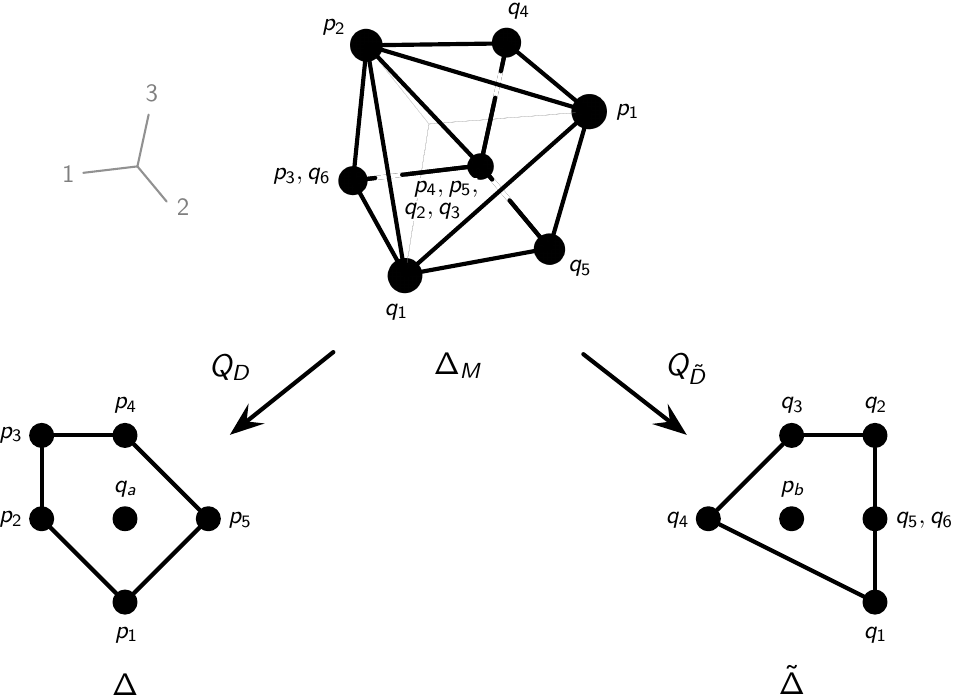}
}
\caption{The matching polytope $\Delta_M$ and the two toric diagrams $\Delta$ and $\tilde{\Delta}$ obtained from it by imposing different D-terms. Every perfect matching corresponds to a distinct point in $\Delta_M$. Since $\Delta_M$ is 6-dimensional in this example, we imposed a 3-dimensional projection to visualize it. The fact that some perfect matchings seem to coincide in $\Delta_M$ is an artifact of this projection. 
\label{fe2}
}
\end{center}
\end{figure}

Every column in this matrix gives the coordinates of a point in $\Delta_M$. The master space is 7-complex dimensional and its toric diagram consists of 11  points. Moreover, the entries in every column of $G_{mast}$ add up to 1, which implies that the master space is CY. Since $\Delta_M$  lives on a hyperplane at distance 1 from the origin, we can project it down to 6 dimensions by dropping one of the rows in \eref{G_mast}. In order to visualize the resulting polytope, we still need to project it to a lower dimension. \fref{fe2} shows one possible projection of $\Delta_M$ to $3d$.

According to \eqref{Q_F}, the kernel of $P$, or equivalently of $G_{mast}$, produces the charge matrix implementing the F-term relations 
\beq
Q_F=
\left(
\begin{array}{ccccc|cccccc}
\ p_1 \ & \ p_2 \ & \ p_3 \ & \ p_4 \ & \ p_5 \ & \ q_1 \ & \ q_2 \ & \ q_3 \ & \ q_4 \ & \ q_5 \ & \ q_6 \
\\
\hline
 1 & 0 & 0 & 1 & 0 & 0 & 0 & 0 & -1 & -1 & 0 \\
 0 & 1 & 0 & 0 & 1 & 0 & 0 & 0 & -1 & 0 & -1 \\
 0 & 0 & 1 & -1 & 1 & 0 & 1 & -1 & 0 & 0 & -1 \\
 0 & 0 & 0 & 0 & 0 & 1 & 1 & 0 & 0 & -1 & -1 \\
\end{array}
\right) \, .
\eeq

The moduli space is the space of solutions to both vanishing F and D-terms. Therefore, it is a projection of the master space onto the subspace of vanishing D-terms. There is a D-term for every gauge group in the theory, with chiral fields contributing to them according to how they transform under the gauge symmetry. Moreover, given the map between GLSM and chiral fields in \eref{map_pm_to_X}, D-terms can be encoded in a charge matrix for GLSM fields, or equivalently perfect matchings, under the gauge groups. We refer to \cite{Franco:2012mm} for details on how this matrix is constructed. 

Gauge groups in the original and twin BFTs correspond to faces and zig-zags of the original brane tiling, respectively. We denote the resulting D-term charge matrices as $Q_D$ and $Q_{\tilde{D}}$. Concatenating each of them with $Q_F$, we obtain the full charge matrices
\beq
Q=\left(\begin{array}{c}Q_F \\ Q_D \end{array}\right) \qquad \qquad \qquad \tilde{Q}=\left(\begin{array}{c}Q_F \\ Q_{\tilde{D}} \end{array}\right)
\eeq

The toric diagrams for the moduli spaces of the original and twin BFTs are then given by
\beq
G=Ker \, Q \qquad \qquad \qquad \tilde{G}=Ker \, \tilde{Q}
\eeq
i.e. the columns in $G$ and $\tilde{G}$ give the positions of the points in $\Delta$ and $\tilde{\Delta}$, respectively.

For the example at hand, we have
{\footnotesize
\beq
Q_D =  
\left(
\begin{array}{ccccc|cccccc}
\ p_1 \ & \ p_2 \ & \ p_3 \ & \ p_4 \ & \ p_5 \ & \ q_1 \ & \ q_2 \ & \ q_3 \ & \ q_4 \ & \ q_5 \ & \ q_6 \
\\
\hline
 0 & 0 & 0 & 0 & 0 & 1 & 0 & 0 & -1 & 0 & 0 \\
 0 & 0 & 0 & 0 & 0 & 0 & 1 & -1 & 0 & 0 & 0 \\
 0 & 0 & 0 & 0 & 0 & 0 & 0 & 1 & 0 & -1 & 0 \\
 0 & 0 & 0 & 0 & 0 & 0 & 0 & 0 & 0 & 1 & -1 \\
\end{array}
\right)
\ \ \
Q_{\tilde{D}} =
\left(
\begin{array}{ccccc|cccccc}
\ p_1 \ & \ p_2 \ & \ p_3 \ & \ p_4 \ & \ p_5 \ & \ q_1 \ & \ q_2 \ & \ q_3 \ & \ q_4 \ & \ q_5 \ & \ q_6 \
\\
\hline
 1 & -1 & 0 & 0 & 0 & 0 & 0 & 0 & 0 & 0 & 0 \\
 0 & 1 & -1 & 0 & 0 & 0 & 0 & 0 & 0 & 0 & 0 \\
 0 & 0 & 1 & -1 & 0 & 0 & 0 & 0 & 0 & 0 & 0 \\
 0 & 0 & 0 & 1 & -1 & 0 & 0 & 0 & 0 & 0 & 0 \\
\end{array}
\right)
\eeq
}
which leads to $\Delta$ and $\tilde{\Delta}$ defined by
{\small
\beq
G =  
\left(
\begin{array}{ccccc|cccccc}
\ p_1 \ & \ p_2 \ & \ p_3 \ & \ p_4 \ & \ p_5 \ & \ q_1 \ & \ q_2 \ & \ q_3 \ & \ q_4 \ & \ q_5 \ & \ q_6 \
\\
\hline
 0 & -1 & -1 & 0 & 1 & 0 & 0 & 0 & 0 & 0 & 0 \\
 -1 & 0 & 1 & 1 & 0 & 0 & 0 & 0 & 0 & 0 & 0 \\
  1 & 1 & 1 & 1 & 1 & 1 & 1 & 1 & 1 & 1 & 1 \\
\end{array}
\right)
\ \ \ 
\tilde{G} =
\left(
\begin{array}{ccccc|cccccc}
\ p_1 \ & \ p_2 \ & \ p_3 \ & \ p_4 \ & \ p_5 \ & \ q_1 \ & \ q_2 \ & \ q_3 \ & \ q_4 \ & \ q_5 \ & \ q_6 \
\\
\hline
 0 & 0 & 0 & 0 & 0 & 1 & 1 & 0 & -1 & 1 & 1 \\
 0 & 0 & 0 & 0 & 0 & -1 & 1 & 1 & 0 & 0 & 0 \\
   1 & 1 & 1 & 1 & 1 & 1 & 1 & 1 & 1 & 1 & 1 \\
\end{array}
\right)
\eeq
}
As for the master space, the entries in every column add up to 1, implying the moduli spaces are CY. This also means that we can drop one row from these matrices, effectively reducing the dimension of the polytope by 1. For these example, $\Delta$ and $\tilde{\Delta}$ turn out to be 2-dimensional, which correspond to CY 3-folds. These two geometries are known as $dP_2$ and $PdP_2$. \fref{fe2} shows the different polytopes involved in this example.

%=================================================================
%=================================================================

In summary $\Delta$ and $\tilde{\Delta}$ arise as two different projections of a single underlying polytope $\Delta_M$ as follows
\beq
\begin{array}{c}
\mbox{{\bf Matching Polytope }} \Delta_M  \\
\begin{array}{ccc} 
\ \ \ \ \swarrow & & \searrow \ \ \ \ \\
\mbox{Gauging faces} & \ \ \ & \mbox{Gauging zig-zags} \\
\downarrow & & \downarrow \\
\Delta & & \tilde{\Delta}
\end{array}
\end{array}
\nonumber
\eeq
where we have identified the gauge groups with the faces and zig-zags of the original brane tiling.

%=================================================================
\section{ Generalized toric polygons}
%=================================================================

\label{section_GTPs}

For the subsequent discussion, it is interesting to enlarge the class of polytopes under consideration. {\it Generalized toric polygons} (GTPs), also known as dot diagrams, were introduced in \cite{Benini:2009gi} to study $(p,q)$ 5-brane webs ending on 7-branes. In particular, GTPs are useful tools for visualizing the $s$-rule, which controls how branes can terminate on others while preserving supersymmetry. We now proceed with a basic review of GTPs and refer the reader to \cite{Benini:2009gi,vanBeest:2020kou,VanBeest:2020kxw} for detailed discussions. GTPs are convex polygons on a $\mathbb{Z}^2$ lattice consisting of two types of points, which we indicate as black and white. GTPs therefore contain and generalize ordinary toric diagrams.

\begin{figure}[H]
\begin{center}
\resizebox{0.5\hsize}{!}{
  \includegraphics[trim=0mm 0mm 0mm 0mm, width=8in]{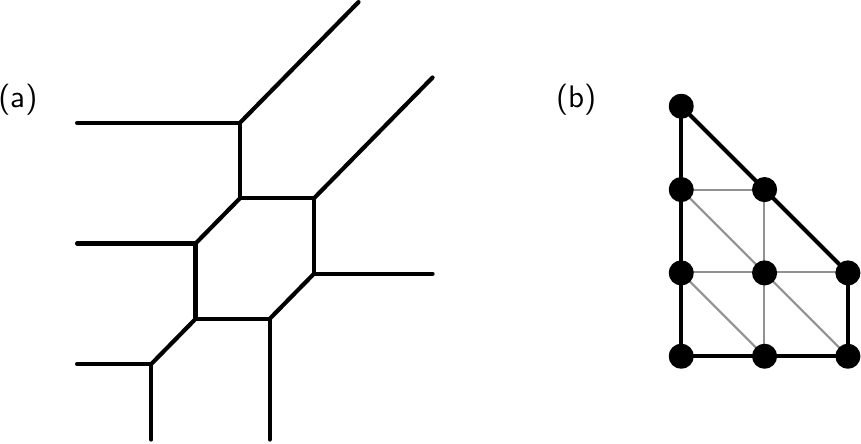}
}
\caption{A $(p,q)$-web and the corresponding triangulated toric diagram. 
\label{pq_web_and_toric_diagram}
}
\end{center}
\end{figure}

Let us first consider $(p,q)$ 5-branes webs without 7-branes, i.e. webs in which the external legs are semi-infinite 5-branes. In this case, the corresponding GTP is a standard toric diagram, i.e. one involving only black dots, obtained from the $(p,q)$-web by graph dualization. \fref{pq_web_and_toric_diagram} shows a $(p,q)$-web and the corresponding toric diagram.

%=================================================================
%=================================================================

Let us now introduce suitable 7-branes to these configurations, such that the legs of the $(p,q)$-web terminate on them. A $(p,q)$ 5-brane can end on a $[p,q]$ 7-brane. Adding 7-is useful for visualizing the Higgs branch of the corresponding gauge theories. The 7-branes can be encoded on the boundary of the GTP as follows. If two parallel legs of the $(p,q)$-web terminate on the same 7-brane, we color the dot that separates the corresponding edges in the polytope in white. Similarly, $n$ consecutive edges on a given side of the GTP separated by $n-1$ white dots represent $n$ parallel legs of the web terminating on a single 7-brane. More generally, a side $\tilde{i}$ of the polytope consisting of $N_{\tilde{i}}$ edges corresponds to $N_{\tilde{i}}$ parallel legs of the web, which can be grouped into sets of $k^{(\tilde{i})}_a$ legs that end on the $a^{th}$ 7-brane. This results into a classification of the corresponding boundary conditions in terms of partitions of $N_{\tilde{i}}$ or, equivalently, Young tableaux with $N_{\tilde{i}}$ boxes. Such configurations translate into different ways of distributing white dots within the corresponding side of the GTP. For a general partition $\{k^{(\tilde{i})}_1,\ldots,k^{(\tilde{i})}_{J_{\tilde{i}}}\}$ into $J_{\tilde{i}}$ sets, with $\sum_{a=1}^{J_{\tilde{i}}} k^{(\tilde{i})}_a=N$, the side of the GTP under consideration splits into sets of $k^{(\tilde{i})}_a$ edges, in which the internal $k^{(\tilde{i})}_a-1$ dots are white. Here $J_{\tilde{i}}$ is the number of 7-branes associated to side $\tilde{i}$ and the $k^{(\tilde{i})}_a$'s are the numbers of parallel 5-branes ending on each of them. Such a side of the GTP contributes an $SU(J_{\tilde{i}})$ factor to the global symmetry of the $5d$ theory engineered by such configuration and of the $4d$ Gaiotto-type theory obtained from it by compactification. In the case of Gaiotto theories, these boundary conditions correspond to different types of punctures. 

\begin{figure}[H]
\begin{center}
\resizebox{0.75\hsize}{!}{
  \includegraphics[trim=0mm 0mm 0mm 0mm, width=8in]{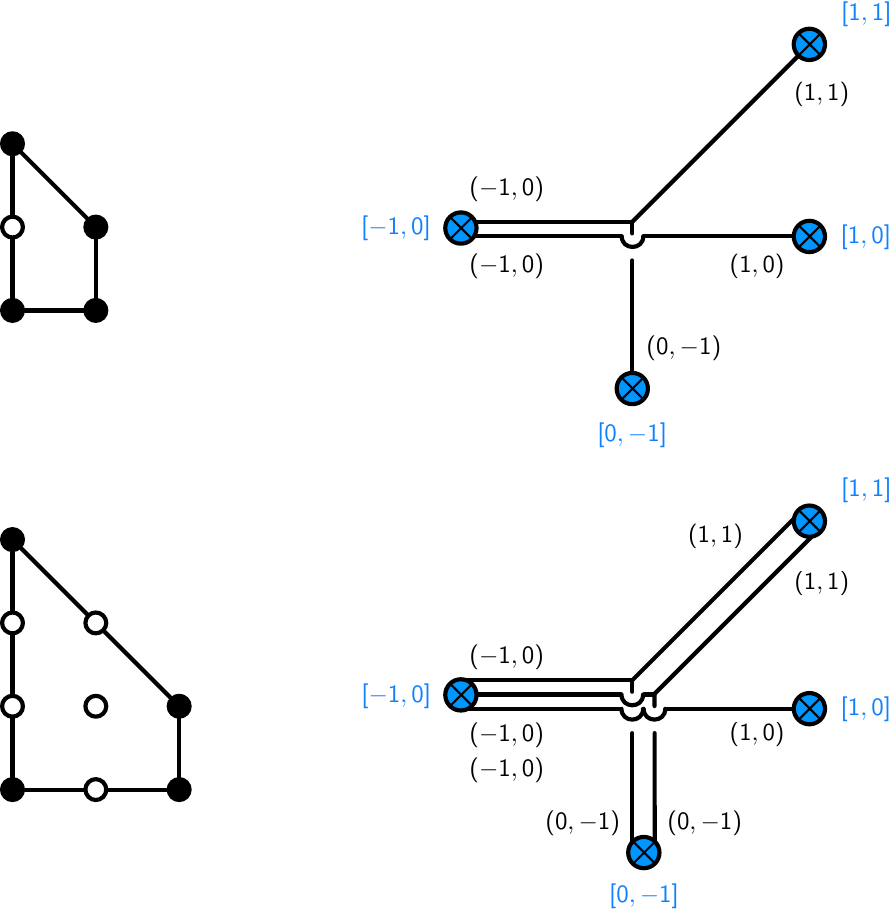}
}
\caption{Two GTPs and the corresponding brane configurations. Blue circles represent 7-branes.
\label{GTP_and_webs}
}
\end{center}
\end{figure}

Finally, once the boundary of the GTP is determined, its interior is constructed by tessellating it with minimal triangles and trapeziums. These building blocks need to obey certain conditions, which can be regarded as the propagation of the $s$-rule into the GTP. This process can result in additional white dots in the interior of the GTP and we refer to \cite{Benini:2009gi} for details. We will return to these conditions in Section \sref{section_s-rule}. \fref{GTP_and_webs} show some examples of GTPs and the dual brane configurations.

%=================================================================
%=================================================================

%=================================================================
\subsection{Polytope mutations and transformations of configurations with 7-branes} 
%=================================================================

\label{section_polytope_mutation_and_7-branes}

\begin{figure}[H]
\begin{center}
\resizebox{0.5\hsize}{!}{
  \includegraphics[trim=0mm 0mm 0mm 0mm, width=8in]{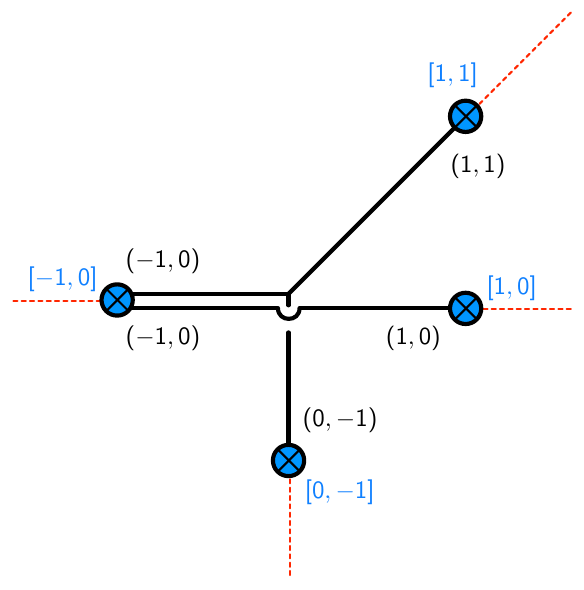}
}
\caption{The first brane configuration in \fref{GTP_and_webs}, now explicitly showing the branch cuts for the 7-branes as dashed red lines.
\label{web_with_branch_cuts}
}
\end{center}
\end{figure}

In the correspondence between polytopes and brane configurations outlined above, polytope mutation is interpreted as a reorganization of the 7-branes. It is often standard to orient the branch cuts of the 7-branes towards infinity, away from the 5-brane web that is suspended from them, as shown in \fref{web_with_branch_cuts} for an example. The 7-branes are thus endowed with a cyclic ordering around the web, which coincides with the ordering of the corresponding sides of the polygon. 

%=================================================================
%=================================================================

The polytope mutation on side $\tilde{j}$ corresponds to the following transformation of the brane setup:
\begin{itemize}
\item The 7-brane $\tilde{j}$ moves to the opposite side of the configuration. This reversal is captured by the first line in \eref{mutation_polytope_1}.
\item The relocation of 7-brane $\tilde{j}$ can be attained by passing over the subset of the 7-branes with either $\langle \tilde{j},\tilde{i} \rangle >0$ or $\langle \tilde{j},\tilde{i} \rangle <0$. In either case, when these 7-branes cross the branch cut of brane $\tilde{j}$, their charges $\eta_{\tilde{i}}$ pick a contribution proportional to $\eta_{\tilde{i}}$ as given by the second line of \eref{mutation_polytope_1}. 
\item The change in rank in \eqref{mutation_polytope_2} amounts to the brane creation effect when the 7-brane $\tilde{j}$ crosses the 5-brane web, and the subsequent brane-antibrane annihilation whenever necessary.
\end{itemize}

\begin{figure}[H]
\begin{center}
\resizebox{0.9\hsize}{!}{
  \includegraphics[trim=0mm 0mm 0mm 0mm, width=8in]{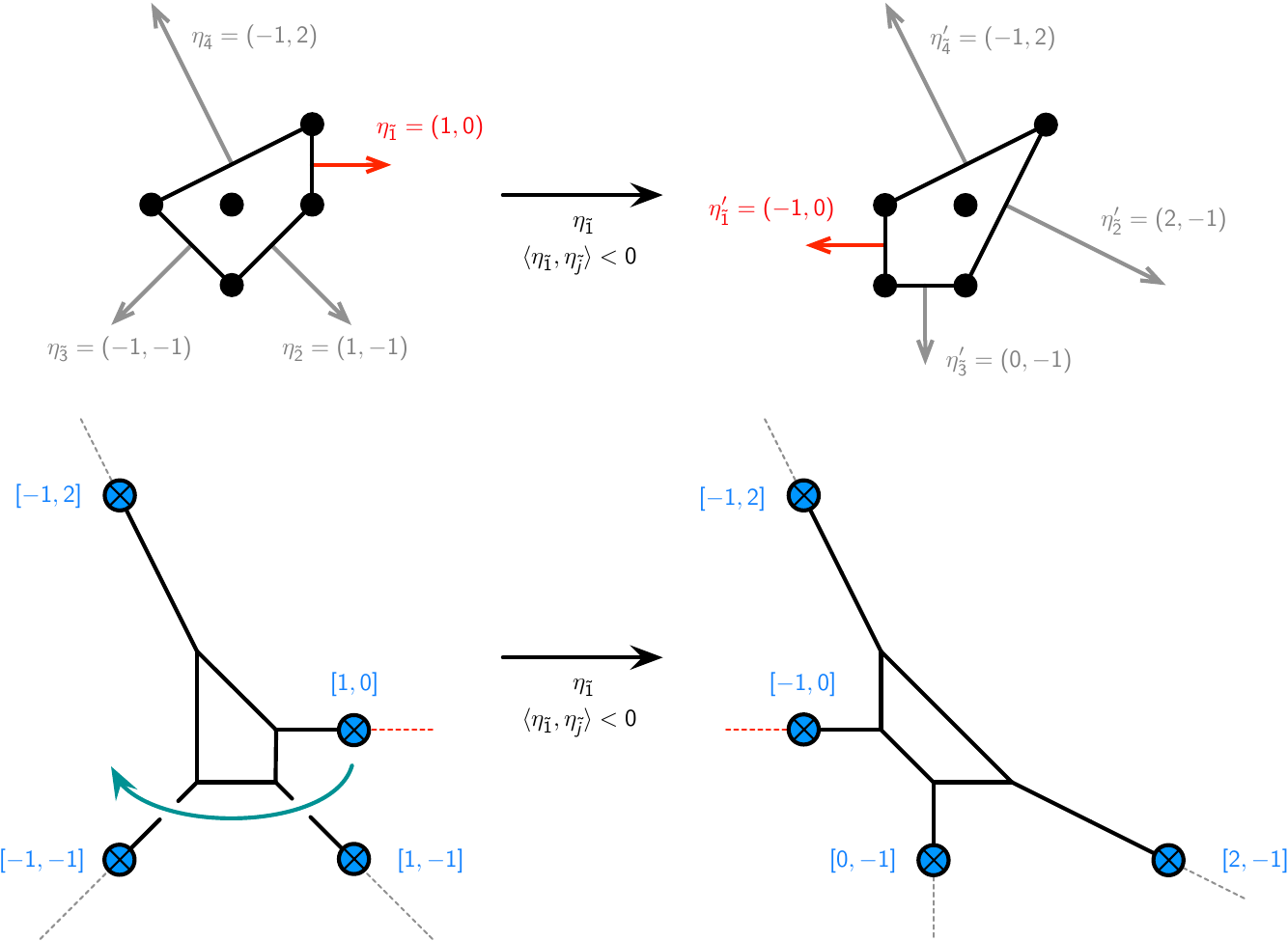}
}
\caption{Polytope mutation and the corresponding transformation of a brane configuration.
\label{polytope_vs_brane_mutation}
}
\end{center}
\end{figure}

\fref{polytope_vs_brane_mutation} shows an example of the brane web transformation associated to a polytope mutation. 

%=================================================================
%=================================================================

%=================================================================
\subsection{Toric diagrams, 5d gauge theories, BPS quivers and beyond} 
%=================================================================

Brane tilings define quiver theories that can be interpreted as either the worldvolume theory on D3-branes probing the toric CY$_3$ \cite{Franco:2005rj} or as BPS quivers encoding the BPS spectrum of the corresponding $5d$ theories \cite{Closset:2019juk}. The generalizations of brane tilings/BPS quivers for GTPs are currently unknown. We expect the ideas presented in this paper are a step towards answering this question.

%=================================================================
\section{Connecting quiver and polytope mutations}
%=================================================================

\label{section_quiver_and_polytope_mutations}

As mentioned in Section \sref{section_democratic_perspective}, the unified treatment of the original and twin realms that we propose raises new questions regarding relations between objects on both sides of the correspondence. With this motivation in mind, in this section we will establish a connection between the mutation of the original polytope at a side $\tilde{i}$, $\mu_{\tilde{i}}(\Delta)$, and the mutation of the twin quiver on the corresponding node $\tilde{i}$, $\mu_{\tilde{i}}(\tilde{Q})$. Due to the symmetry between the original and twin sides, the same argument connects $\mu_{i}(\tilde{\Delta})$ to the mutation of the original quiver $\mu_{i}(Q)$. 

The alert reader might recognize that the transformation in \eref{mutation_polytope_1} and \eref{mutation_polytope_2} defining the polytope mutation, often appears in geometric realizations of Seiberg duality/quiver mutation, where the $\eta_{\tilde{i}}$ represent brane charges (see e.g. \cite{Cachazo:2001sg}). Generally speaking, in all these constructions, the dualization of the gauge group associated to the branes of type $\eta_{\tilde{j}}$ is attained by inverting $\eta_{\tilde{j}}$ and ``moving itover " all the branes with positive intersection with it, i.e. $\langle \eta_{\tilde{j}},\eta_{\tilde{i}}\rangle > 0$.\footnote{In \cite{Feng:2002kk}, the interesting possibility of moving $\eta_{\tilde{j}}$ over only a subset of the branes with $\langle \eta_{\tilde{j}},\eta_{\tilde{i}}\rangle > 0$ was suggested. This operation was denoted {\it fractional Seiberg duality}. However, based on what happens if we partially realize the brane reorganization associated to Seiberg duality in other simple setups, such as Hanany-Witten configurations \cite{Elitzur:1997fh}, it is natural to expect that such configurations would break SUSY.} From the point of view of quivers, those are the nodes at the end of arrows coming out of node $\tilde{j}$.\footnote{As in polytope mutation, we can alternatively move the branes with $\langle \eta_{\tilde{j}},\eta_{\tilde{i}}\rangle < 0$, which correspond to nodes in the quiver from which arrows go into node $\tilde{j}$.} Equation \eref{mutation_polytope_1} summarizes the transformation of brane charges in this process and \eref{mutation_polytope_2} states how the multiplicity of branes of type $\tilde{j}$ needs to change in order to preserve the total brane charge. 

But what is the quiver whose mutation corresponds to polytope mutation? Let us first consider the case in which all $N_{\tilde{i}}=1$, namely a polytope in which all sides have a single edge. Clearly, this transformation is the mutation of a quiver whose nodes correspond to zig-zag paths. But this quiver is precisely the theory $\tilde{Q}$ obtained from the original one by untwisting. We therefore conclude that the mutation of the original polytope corresponds to the mutation of the twin quiver. Loosely speaking,
\beq
\mu_{\tilde{i}}(\Delta) = \mu_{\tilde{i}}(\tilde{Q}) \, .
\eeq

\begin{figure}[H]
\begin{center}
\resizebox{0.65\hsize}{!}{
  \includegraphics[trim=0mm 0mm 0mm 0mm, width=8in]{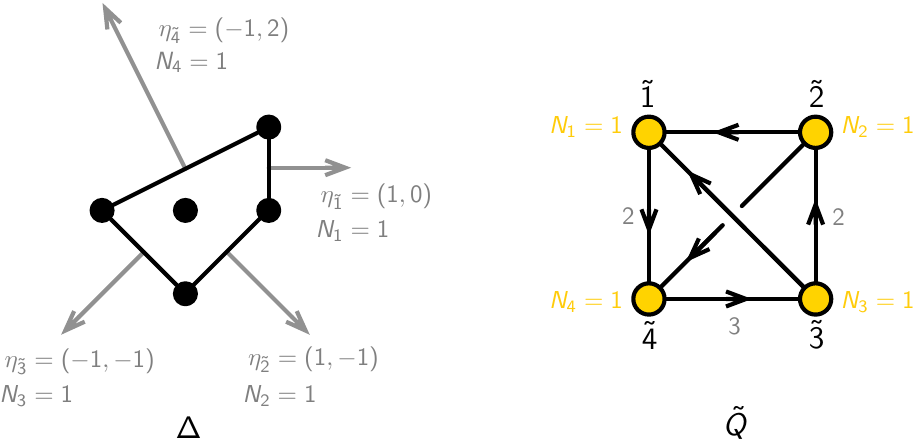}
}
\caption{Toric diagram for $dP_1$ and the corresponding twin quiver. This case is special in that the twin and original quivers coincide.
\label{dP1_Delta_and_twin_quiver}
}
\end{center}
\end{figure}

To illustrate these ideas, let us revisit the example discussed in Section \sref{section_polytope_mutation}. The original polytope shown in \fref{dP1_Delta_and_twin_quiver} is $dP_1$, whose quiver theory (equivalently dimer model) has been studied at length \cite{Feng:2000mi,Feng:2001xr,Feng:2002zw,Franco:2005rj}. This theory is rather special, since the twin theory obtained by untwisting coincides with the original one, namely $\tilde{Q}=Q$ and $\tilde{\Delta}=\Delta$ \cite{Hanany:2012vc}.\footnote{For all regular reflexive polytopes, there is at least one toric phase with the property that the original and twin theories coincide. However, this feature is not generic. For example, it clearly cannot hold for polytopes in which the area and perimeter are not equal, since this corresponds to a different number of faces and zig-zag paths in the corresponding BFT and, consequently, different numbers of nodes in the original and twin quivers. This special behavior is interesting in its own right, but it will not play any role in our subsequent discussion. It merely simplifies our presentation, since everything reduces to a single theory that has been thoroughly studied in the literature.} 

%=================================================================
%=================================================================

We see that there are two qualitative different types of nodes in the quiver. Nodes 1 and 2, which have two incoming (outgoing) arrows and hence $N_f=2 N_c$. Such nodes are often referred to as {\it toric nodes} in the literature. Starting from a toric phase and dualizing a toric node results in a new toric phase. On the other hand, Nodes 3 and 4 have more than two incoming (outgoing) arrows and hence $N_f>2 N_c$.\footnote{Formally, we can also act with quiver mutations on nodes with $N_f=N_c$ or $N_f<N_c$. In the first case, the rank of the mutated node is zero, which corresponds to the disappearance of the associated side of $\Delta$. The second case, in which the rank of the mutated node becomes negative, will be considered in Section \sref{section_s-rule}.} In fact, the number of arrows between two nodes $\tilde{i}$ and $\tilde{j}$ in the quiver $\tilde{Q}$ can be directly computed from the toric diagram as the intersection number $\langle \eta_{\tilde{i}}, \eta_{\tilde{j}} \rangle$ defined in \eref{intersection_sides}, where the sign determines the orientation.\footnote{We will elaborate on this in Section \sref{section_direction_construction_quiver_for_GTP}. These intersection numbers are insensitive to the symmetric part of the adjacency matrix, namely to vector-like pairs of arrows connecting two nodes in opposite directions.} Starting from a toric phase and dualizing a non-toric node results in a non-toric phase. Below we consider the mutation on each type of node and how the transformation of the quiver correlates with what happen to the polytope.

%=================================================================
\subsection*{Mutation of $\tilde{Q}$ on a toric node}
%=================================================================

\begin{figure}[H]
\begin{center}
\resizebox{0.65\hsize}{!}{
  \includegraphics[trim=0mm 0mm 0mm 0mm, width=8in]{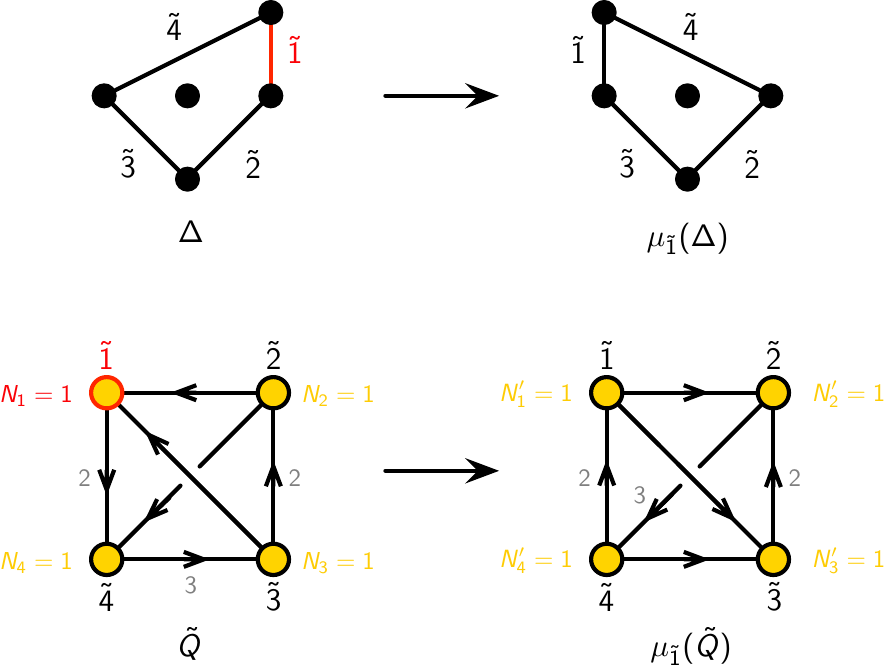}
}
\caption{Mutation of $\tilde{Q}$ on a toric node and corresponding mutation of $\Delta$.
\label{mutation_dP1_on_toric_node}
}
\end{center}
\end{figure}

Let us mutate $\tilde{Q}$ on node $\tilde{1}$ (mutating node $\tilde{2}$ is equivalent) and, correspondingly, the polytope $\Delta$ on side $\tilde{1}$. \fref{mutation_dP1_on_toric_node} shows the effect of the polytope and quiver mutations.

%=================================================================
%=================================================================

As expected, the mutated twin quiver is again a toric phase. In fact, this example is even more special, since $\mu_{\tilde{1}}(\tilde{Q})$ is identical to $\tilde{Q}$ up to a relabeling of the nodes. This is nicely mirrored by the fact that the mutated polytope is equal to the original one. More generally, the mutation of the twin quiver on a toric node might result on a different toric phase. In such a case, the mutated polytope $\mu_{\tilde{i}}(\Delta)$ would be different from $\Delta$,\footnote{We consider $n$-dimensional polytopes up to $SL(n,\mathbb{Z})$ equivalences.} but every side would still have $N_{\tilde{i}}'=1$, corresponding to a toric phase.  

Notice that we applied quiver mutation to a node with rank 1. If we want to interpret the mutation as Seiberg duality, nodes need to be non-abelian. This is simply addressed by interpreting the $\tilde{N}_i$’s as giving ranks only up to an overall common factor $N$. Alternatively, we can simply regard the quiver mutation as a formal operation.

%=================================================================
\subsection*{Mutation of $\tilde{Q}$ on a non-toric node}
%=================================================================

\begin{figure}[H]
\begin{center}
\resizebox{0.7\hsize}{!}{
  \includegraphics[trim=0mm 0mm 0mm 0mm, width=8in]{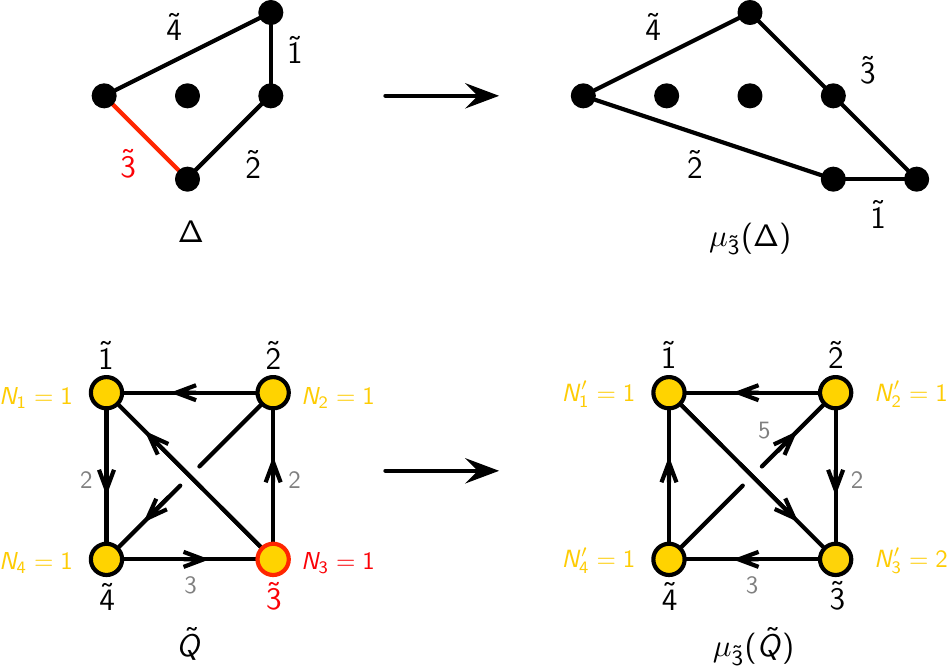}
}
\caption{Mutation of $\tilde{Q}$ on a non-toric node and corresponding mutation of $\Delta$.
\label{mutation_dP1_on_non-toric_node}
}
\end{center}
\end{figure}

Let us now mutate $\tilde{Q}$ on node $\tilde{3}$ (mutating node $\tilde{4}$ is equivalent) and, correspondingly, the polytope $\Delta$ on side $\tilde{3}$. \fref{mutation_dP1_on_non-toric_node} shows the effect of the polytope and quiver mutations.

%=================================================================
%=================================================================

In this case, the rank of the dualized node becomes $N'_{\tilde{3}}$=2 and therefore the mutated twin quiver is a non-toric phase. Accordingly, the mutated polytope $\mu_{\tilde{3}}(\Delta)$ exhibits a side of length 2. Of course, the new quiver can also be determined from the polytope $\mu_{\tilde{3}}(\Delta)$ by computing the new intersection numbers $\langle \eta'_{\tilde{i}}, \eta'_{\tilde{j}} \rangle$.

\bigskip

It is worth noting that a discussion similar to the one in this section can be found in \cite{Feng:2002kk}. That work investigated the realization of Seiberg duality on theories on D3-branes probing toric CY 3-folds as Picard-Lefschetz transformations in the mirror geometry, where they are related to mutations of $(p,q)$-webs (equivalently polytopes). The main difference between those references and our work is that those earlier papers did not carefully take into account the distinction between the two sides of the correspondence, namely between $Q$ and $\tilde{Q}$ or $\Delta$ and $\tilde{\Delta}$. One reason for this is that \cite{Feng:2002kk} primarily focused on reflexive polytopes for which, as mentioned above, there is always a phase for which $Q$ and $\tilde{Q}$ coincide. Moreover, the concept of untwisting was not available in the pre-brane tiling era. In the language of this paper, the work in \cite{Feng:2002kk} can be regarded as studying the equivalent problem of quiver mutations $\mu_i(Q)$ as mutations on the twin polytope $\mu_i(\tilde{\Delta})$.

Interestingly, some earlier works incorrectly emphasized the apparent independence of quiver and polytope mutations, see e.g. \cite{Higashitani:2019vzu}. We have just shown that they are indeed intimately related. The key point is to think in terms of twin quivers, instead of the original ones.

%=================================================================
\section{Twin quivers for toric diagrams and GTPs}
%=================================================================

\label{section_twin_quivers_for_GTPs}

Motivated by the correspondence between polytope and quiver mutations, we observed in the previous section that non-toric twin quivers $\tilde{Q}$ are associated with polytopes $\Delta$ for which some of the sides have more than one edge, i.e. $N_{\tilde{i}} > 1$. However, it is still necessary to think more carefully about the precise interpretation of this polytope. We will now argue that the natural interpretation of the mutated polytope associated to a non-toric $\tilde{Q}$ is not as a toric diagram, but as a GTP.

We will illustrate our ideas with the example in the top right of \fref{mutation_dP1_on_non-toric_node}. Let us forget that this polytope was obtained by mutation and simply call it $\Delta$. We will alternatively interpret the polytope as a toric diagram and as certain GTP, and conclude that a non-toric $\tilde{Q}$ captures properties of the latter.

%=================================================================
\subsection{$\Delta$ as a toric diagram}
%=================================================================

\label{Delta_as_toric_diagram}

\begin{figure}[H]
\begin{center}
\resizebox{0.8\hsize}{!}{
  \includegraphics[trim=0mm 0mm 0mm 0mm, width=8in]{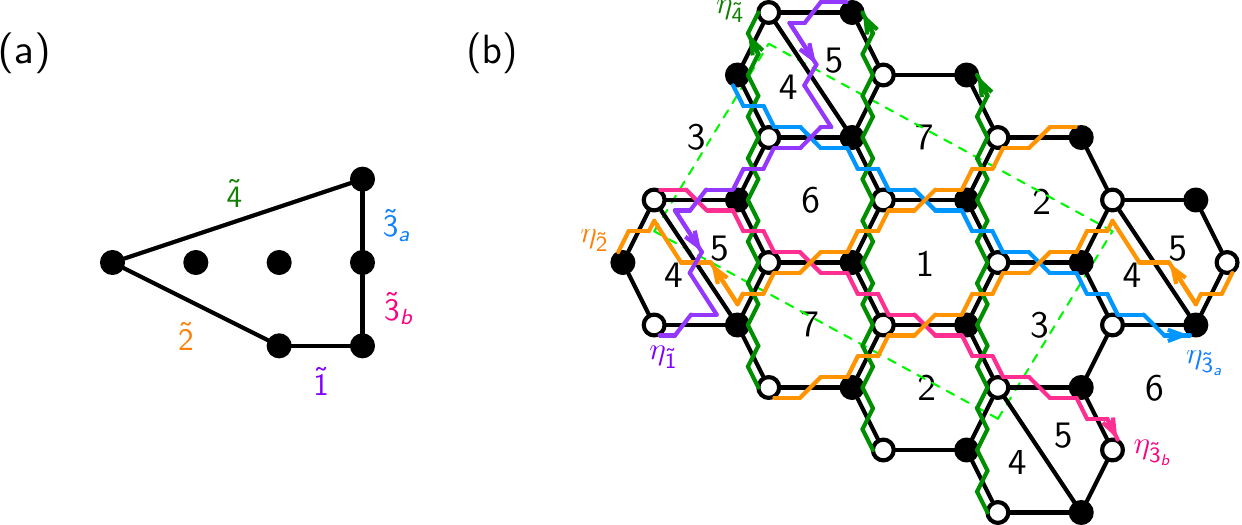}
}
\caption{Toric diagram $\Delta$ for $X^{3,3}$ and the brane tiling for the corresponding quiver theory $Q$.
\label{fe07}
}
\end{center}
\end{figure}

Let us first think about $\Delta$ as a toric diagram, as shown in \fref{fe07}.a. We have applied an $SL(2,\mathbb{Z})$ transformation to take the toric diagram to a standard form. It is straightforward to see that interpreting the polytope as a toric diagram contradicts the discussion in the previous section, whose salient feature is that it leads to a beautiful correspondence between polytope and quiver mutations. We can construct a brane tiling for $\Delta$ regarded as a toric diagram or, equivalently, the corresponding theory $Q$. There are several efficient methods for doing so, see e.g. \cite{Franco:2005rj,Hanany:2005ss,Franco:2012mm}. The toric diagram under consideration corresponds to $X^{3,3}$ in the classification of \cite{Hanany:2005hq}, where the corresponding quiver theory was explicitly constructed. We can simply borrow these results to construct the brane tiling, which is shown in \fref{fe07}.\footnote{We constructed the toric phase with a minimal number of chiral fields. In Section \sref{section_twin_quivers_for_different_toric_phases}, we will discuss the twin quivers associated to different toric phases.} $\Delta$ has four sides, but one of them contains two segments, so its perimeter consists of five segments. This implies that the brane tiling for $Q$ has five zig-zag paths, as explicitly shown in \fref{fe07}. Interestingly, we can conclude this directly from $\Delta$, even without constructing $Q$. Correspondingly, we know in advance that in this case $\tilde{Q}$ has five nodes, each of them with $N_{\tilde{i}}=1$. 

%=================================================================
%=================================================================

\begin{figure}[H]
\begin{center}
\resizebox{0.9\hsize}{!}{
  \includegraphics[trim=0mm 0mm 0mm 0mm, width=8in]{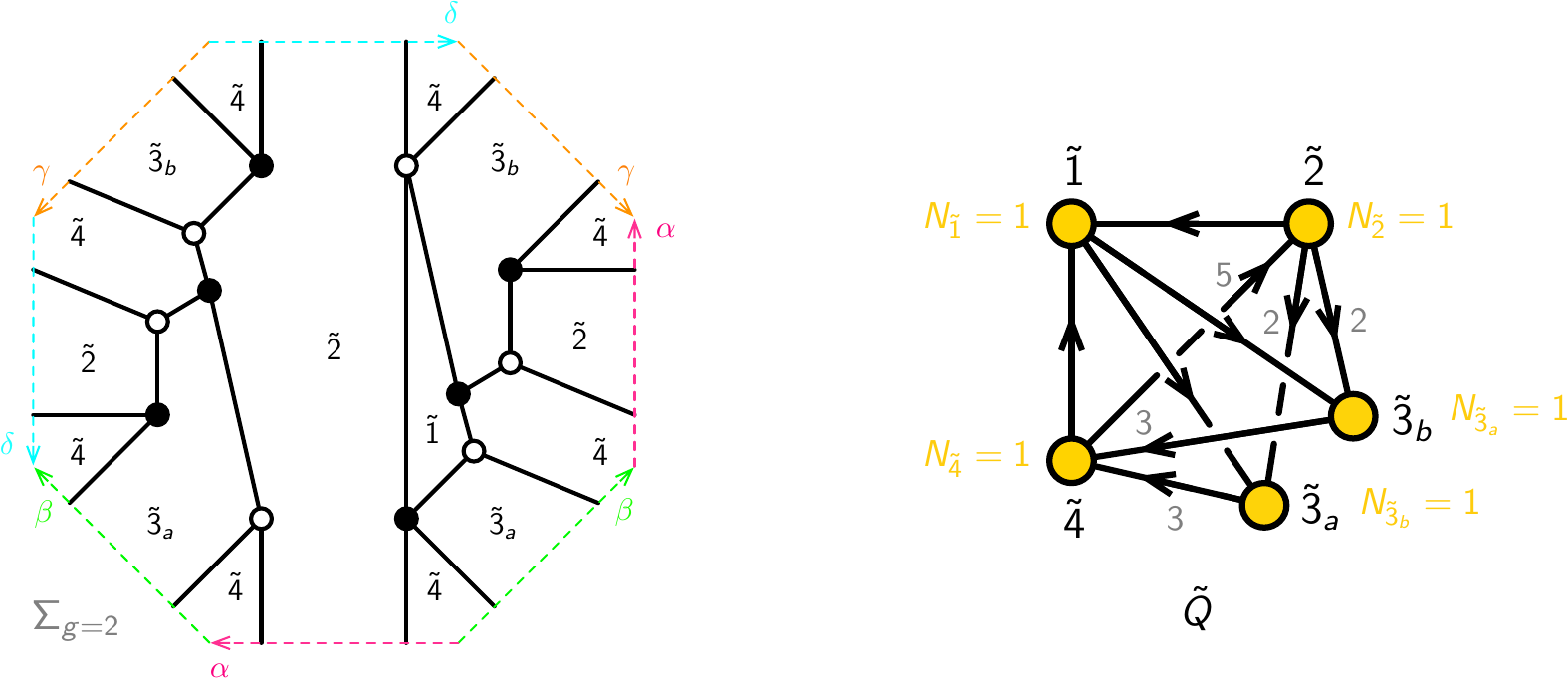}
}
\caption{The BFT $\tilde{Q}$ obtained by untwisting the the brane tiling in \fref{fe07}. The bipartite graph is now embedded on a genus 2 Riemann surface, as expected from $\Delta$. We show the fundamental domain and how the segments on its boundary are identified. On the right, we show the corresponding quiver. 
\label{fe08}
}
\end{center}
\end{figure}

Once we have the brane tiling for $Q$, the BFT for $\tilde{Q}$ is constructed by untwisting, as shown in \fref{fe08}. This procedure not only produces the twin quiver, but also its superpotential. As expected, $\tilde{Q}$ has five $N_{\tilde{i}}=1$ nodes. Clearly, this is not the non-toric quiver in \fref{mutation_dP1_on_non-toric_node}. However, we will see in Section \sref{section_connection_toric_GTP} that the two are related.

%=================================================================
%=================================================================

An interesting algorithm for deriving brane tilings for mutated polytopes, regarding them as toric diagrams, was introduced in \cite{Higashitani:2019vzu}. We will not explain the details of this construction here and instead refer the reader to the original paper. As we mentioned earlier, there are many ways of deriving a brane tiling directly from the new toric diagram. However, this algorithm is special in that it connects the brane tiling for the initial toric diagram to a specific new brane tiling determined by the mutation. We will use this feature in the analysis of Section \sref{section_twin_quivers_for_different_toric_phases}.

%=================================================================
\subsection{$\Delta$ as a GTP}
%=================================================================

\label{section_Delta_as_a_GTP}

\begin{figure}[H]
\begin{center}
\resizebox{0.8\hsize}{!}{
  \includegraphics[trim=0mm 0mm 0mm 0mm, width=8in]{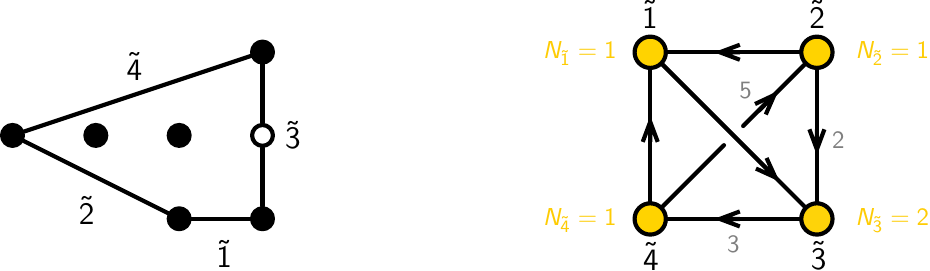}
}
\caption{We proposed that the twin quiver $\tilde{Q}$ associated to a GTP with white does is a non-toric phase. Therefore, the polytope obtained in \fref{mutation_dP1_on_non-toric_node} by mutation should be interpreted as a GTP.
\label{fe09}
}
\end{center}
\end{figure}

We propose that the non-toric twin quiver is naturally associated to a GTP, as shown in \fref{fe09}. This correspondence is partly motivated by the facts that $\tilde{Q}$ has a node for each side of the polygon, the number of segments in each side correspond to their ranks, and polytope and quiver mutations are elegantly unified in this interpretation. In coming sections, we will argue that global symmetries and the $s$-rule also support this interpretation. While we will not discuss its superpotential in further detail, when the non-toric phase is obtained by mutation, it can be determined from the original theory by the standard rules of Seiberg duality. 

%=================================================================
%=================================================================

While we arrived to the correspondence between GTPs and non-toric twin quivers via mutations, we propose that it holds in general. Namely, that toric diagrams and GTPs correspond to toric and non-toric twin quivers, respectively. In Section \sref{section_general_boundary_conditions}, we will elaborate on the twin quivers for more general GTPs, in which not all the internal dots in a side are white.

%=================================================================
\subsection{Connection between the toric diagram and the GTP}
%=================================================================

\label{section_connection_toric_GTP}

\begin{figure}[H]
\begin{center}
\resizebox{0.8\hsize}{!}{
  \includegraphics[trim=0mm 0mm 0mm 0mm, width=8in]{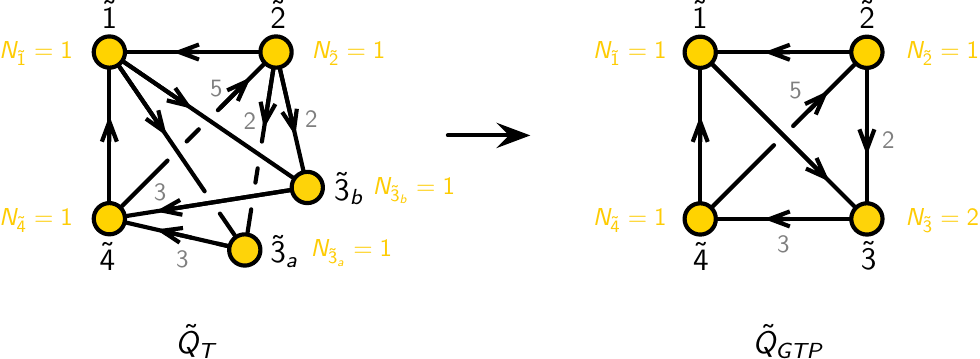}
}
\caption{Starting from $\tilde{Q}_{\mbox{T}}$, we obtain $\tilde{Q}_{\mbox{GTP}}$ by combining nodes $\tilde{3}_a$ and $\tilde{3}_b$ into a single rank-2 node $\tilde{3}$.\label{QT_to_QGTP}
}
\end{center}
\end{figure}

To simplify the discussion in this subsection, we will refer to the twin quivers for the toric diagram and GTP as $\tilde{Q}_{\mbox{T}}$ and $\tilde{Q}_{\mbox{GTP}}$, respectively. The previous examples show that both twin quivers are related in a simple way, which can be summarized as follows:

\begin{itemize}
\item $\tilde{Q}_{\mbox{T}} \to \tilde{Q}_{\mbox{GTP}}$: combine $N_{\tilde{i}}$ rank-1 nodes into a single rank-$N_{\tilde{i}}$ node.
\item $\tilde{Q}_{\mbox{GTP}} \to \tilde{Q}_{\mbox{T}}$: split a rank-$N_{\tilde{i}}$ node into $N_{\tilde{i}}$ rank-1 nodes.
\end{itemize}
\medskip
\fref{QT_to_QGTP} illustrates this process for the two twin quivers discussed in the previous Sections \ref{Delta_as_toric_diagram} and \ref{section_Delta_as_a_GTP}. We can think about the process taking from $\tilde{Q}_{\mbox{GTP}}$ to $\tilde{Q}_{\mbox{T}}$ as adjoint higgsing.\footnote{The field in the adjoint representation of node $\tilde{i}$ that would be responsible for such higgsing is not part of the quiver $\tilde{Q}_{\mbox{GTP}}$.}   

%=================================================================
%=================================================================

This connection provides an alternative algorithm for generating $\tilde{Q}_{\mbox{GTP}}$ that does not rely on mutations:

\begin{enumerate}
\item Regarding the polytope under consideration as a toric diagram, we construct the corresponding brane tiling. 
\item Acting on it with untwisting, we obtain the BFT for $\tilde{Q}_{\mbox{T}}$, which contains one node for every zig-zag path of the original tiling/edge of the original toric diagram. 
\item Finally, the $N_{\tilde{i}}$ nodes associated to every side with $N_{\tilde{i}}$ edges in the original toric diagram are merged into a single node of rank $N_{\tilde{i}}$. 
\end{enumerate}

%=================================================================
\section{Direct construction of the twin quiver for a GTP}
%=================================================================

\label{section_direction_construction_quiver_for_GTP}

It is possible to construct twin quivers directly from GTPs as follows:

\begin{enumerate}
\item For every side $\tilde{i}$ of the GTP introduce a rank-$N_{\tilde{i}}$ node.
\item The arrows connecting every pair of nodes $\tilde{i}$ and $\tilde{j}$ are given by the intersection number in \eref{intersection_sides}, i.e. 
\beq
A_{\tilde{i} \tilde{j}}=\langle \eta_{\tilde{i}},\eta_{\tilde{j}} \rangle = \det \left(\begin{array}{cc} p_{\tilde{i}} & q_{\tilde{i}} \\ p_{\tilde{j}} & q_{\tilde{j}} \end{array}\right) \, ,
\eeq
where the sign of $A_{\tilde{i} \tilde{j}}$ determines the orientation of the corresponding arrow(s). In our previous discussions, we have adopted the convention in which $A_{\tilde{i} \tilde{j}}>0$ corresponds to $\tilde{i} \to \tilde{j}$ arrows and $A_{\tilde{i} \tilde{j}}<0$ corresponds to $\tilde{j} \to \tilde{i}$ arrows.
\end{enumerate}

Strictly speaking, this procedure is only sensitive to the antisymmetric part of the adjacency matrix and therefore cannot detect the possible presence of bidirectional arrows between nodes or adjoint fields. We will revisit this issue in Section \sref{section_twin_quivers_for_different_toric_phases}. In addition, it does not produce the superpotential for $\tilde{Q}$. When $\tilde{Q}$ is a toric phase, the superpotential can be obtained by deriving the full BFT by untwisting. For non-toric $\tilde{Q}$, the superpotential can be determined if the theory can be connected to a BFT by quiver mutations. It is therefore reasonable to expect that a general method for deriving the superpotential from a GTP without relying on mutations exists. We leave this interesting question for future work.

%=================================================================
\subsection{Twin quivers for general boundary conditions}
%=================================================================

\label{section_general_boundary_conditions}

So far we have discussed GTPs in which all dots on a side other than the corners are white. This corresponds to all the 5-branes associated those edges terminating on a single D7-brane. The discussion in the previous sections leads to a natural candidate for the twin quivers corresponding to the more general boundary conditions mentioned in Section \sref{section_GTPs}, which are represented by GTPs with more general arrangements of white and black dots on their boundary.  For each of side, we have a partition of $N_{\tilde{i}}$, $\{k^{(\tilde{i})}_1,\ldots,k^{(\tilde{i})}_{J_{\tilde{i}}}\}$.

In this case, we propose that $\tilde{Q}$ is given by the straightforward generalization of the method in Section \sref{section_direction_construction_quiver_for_GTP}:
\begin{enumerate}
\item For every side $\tilde{i}$ of the GTP introduce $J_{\tilde{i}}$ nodes, each of them with rank $k^{(\tilde{i})}_a$.
\item  Every node in the set $\tilde{i}$ is connected to every node in the set $\tilde{j}$ by arrows given by the intersection number $\langle \eta_{\tilde{i}},\eta_{\tilde{j}} \rangle$.
\end{enumerate}

We can also construct $\tilde{Q}$ for general boundary conditions with a similar generalization with the procedure in Section \sref{section_connection_toric_GTP}, namely:
\begin{enumerate}
\item Regarding the polytope under consideration as a toric diagram, construct the corresponding brane tiling. 
\item Generate the BFT for $\tilde{Q}_{\mbox{T}}$ by untwisting. This quiver has one node for every zig-zag path of the original tiling/edge of the original toric diagram. 
\item Finally, the $N_{\tilde{i}}$ nodes associated to every side with $N_{\tilde{i}}$ edges in the original toric diagram are merged into $J_{\tilde{i}}$ nodes of rank $k^{(\tilde{i})}_1,\ldots,k^{(\tilde{i})}_{J_{\tilde{i}}}$. 
\end{enumerate}

\begin{figure}[H]
\begin{center}
\resizebox{0.8\hsize}{!}{
  \includegraphics[trim=0mm 0mm 0mm 0mm, width=6in]{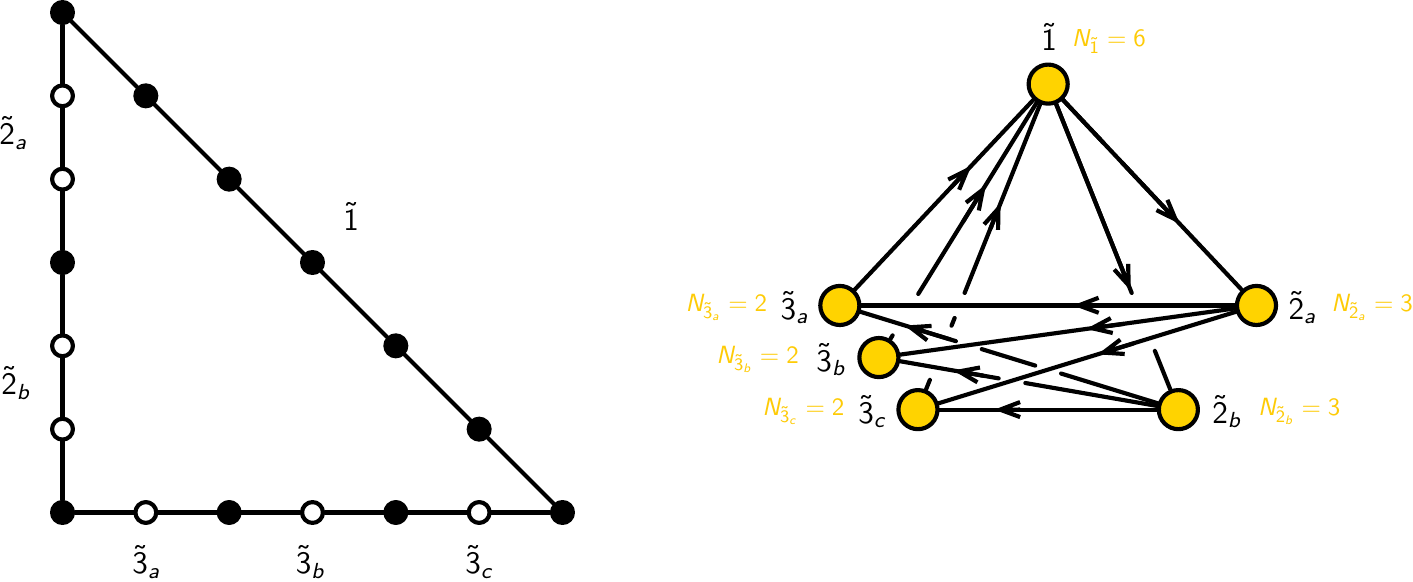}
}
\caption{A GTP with general boundary conditions, for which we only show the boundary, and the corresponding twin quiver.
\label{example_general_boundary_conditions}
}
\end{center}
\end{figure}

\fref{example_general_boundary_conditions} shows an example of this construction.

%=================================================================
%=================================================================

%=================================================================
\paragraph{Global symmetry.} 
%=================================================================

Twin quivers reflect the global symmetry of the $4d$ theories associated to GTPs. These symmetries arise as permutations of nodes in the twin quiver. The simplest example corresponds to the so called {\it full} or {\it maximal punctures}. A theory in which all punctures are of this type correspond to a 7-brane for each segment on the boundary of the polytope, namely to $J_{\tilde{i}}=N_{\tilde{i}}$ for all $\tilde{i}$. In this case the GTP is a standard toric diagram, there are $N_{\tilde{i}}$ nodes for each side $\tilde{i}$ of the GTP and the quiver exhibits an $S_{ N_{\tilde{i}}}$ symmetry that permutes them. This permutation group is the Weyl group of the $SU(N_{\tilde{i}})$ global symmetry that arises when the corresponding 7-branes coincide.

%=================================================================
\section{Twin quivers from different toric phases}
%=================================================================

\label{section_twin_quivers_for_different_toric_phases} 

In previous sections, to simplify our presentation, we often talked about {\it the} brane tiling or BFT $Q$ for an original toric diagram $\Delta$. However, generically, a given toric diagram is associated to multiple BFTs. These different theories are known as toric phases and are connected to each other by mutations on toric nodes. Below, we investigate how the non-uniqueness of $Q$ reflects on the twin quivers $\tilde{Q}$, focusing on the generic case of GTPs. This issue impacts the different methods that we introduced for deriving twin quivers for GTPs.

We will illustrate the discussion using explicit examples. In each case, we will construct $\tilde{Q}$ for a GTP in two independent ways, via mutation as in Section \sref{section_quiver_and_polytope_mutations} and using the method in Section \sref{section_connection_toric_GTP}. The method in Section \sref{section_direction_construction_quiver_for_GTP} does not use an original quiver theory $Q$ as a starting point, so its result cannot depend on toric phases. Our discussion will clarify how this apparent discrepancy is accounted for by the known limitations of this method.

\begin{figure}[ht!]
\begin{center}
\resizebox{0.7\hsize}{!}{
  \includegraphics[trim=0mm 0mm 0mm 0mm, width=9in]{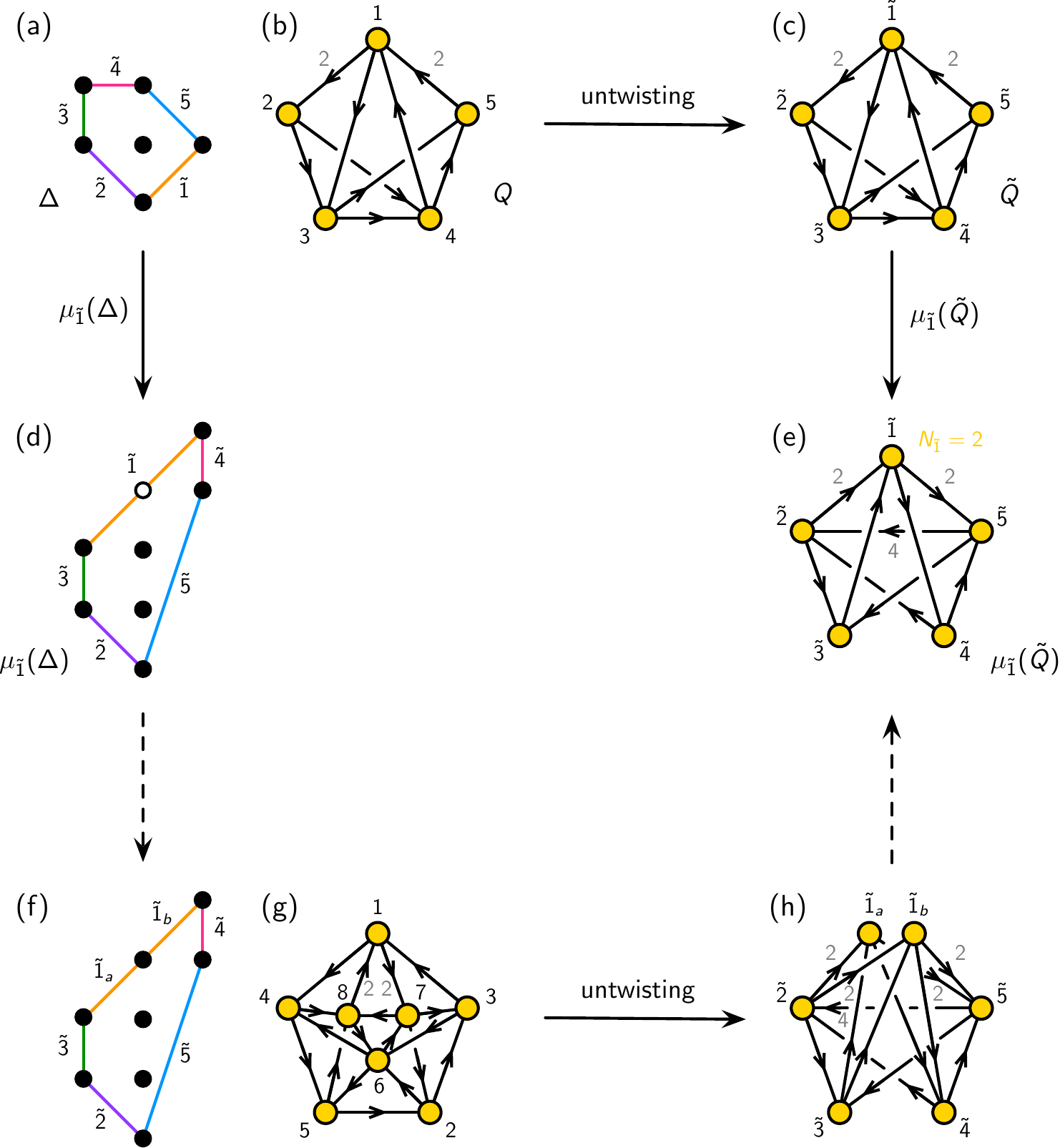}
 }
\caption{Derivation of the twin quiver for the GTP in (d) via two methods, using $dP_{2,(a)}$ as a starting point. The top row uses the algorithm presented in Section \sref{section_quiver_and_polytope_mutations}, while the bottom row uses the one in Section \sref{section_connection_toric_GTP}. Both methods produce the same twin quiver, which is shown in (e).
\label{f_dP2a_mutation2}
}
\end{center}
\end{figure}

All the explanation that follows refers to subfigures in \fref{f_dP2a_mutation2}. Let us consider $dP_2$, whose toric diagram is shown in $(a)$. This theory has been studied at length in the literature and has two toric phases, which we will denote $dP_{2,(a)}$ and $dP_{2,(b)}$ \cite{Feng:2001xr,Feng:2002zw,Franco:2005rj}. Additional details about the models considered in this section, including graphs before and after untwisting, are presented in Appendix \sref{appendix_details_models}. Let us start from $dP_{2,(a)}$, as shown in $(b)$. Mutating the toric diagram on side $\tilde{1}$ results in the GTP in $(d)$. We are interested in finding the corresponding twin quiver. The first approach is summarized on the top row of the figure. Acting with untwisting on $dP_{2,(a)}$  we obtain the twin quiver in $(c)$. Mutating this quiver on node $\tilde{1}$ produces the theory in $(e)$, which is the twin quiver for the GTP we are interested in. The mutation also produces the superpotential for this theory. Notice that, as in previous examples, we do not assign an original quiver $Q$ to the GTP. This is an important open question that we revisit in the conclusions.

Let us derive the same theory with a different procedure, shown on the bottom row of \fref{f_dP2a_mutation2}. This time, we first trade the GTP for a toric diagram, as in $(f)$. We then build a brane tiling for it, for which we can use several methods. This particular toric diagram has multiple toric phases, i.e. multiple brane tilings associated to it. In order to single out which toric phase to consider, we used the algorithm of \cite{Higashitani:2019vzu}. Without going into details, this construction generates a specific phase of the toric diagram, which is determined by the original phase, in this case $dP_{2,(a)}$, and the mutation of the polytope under consideration. The result is shown in $(g)$.  We then untwist this phase to obtain $(h)$. Finally we combine nodes $\tilde{1}_a$ and $\tilde{1}_b$, which correspond to the two edges on side $\tilde{1}$, into a rank-2 node. The result is again the quiver in $(e)$. The fact that we obtained the same result in two different ways is a nice test of the proposed algorithms. 

%=================================================================
%=================================================================

\begin{figure}[ht!]
\begin{center}
\resizebox{0.7\hsize}{!}{
  \includegraphics[trim=0mm 0mm 0mm 0mm, width=9in]{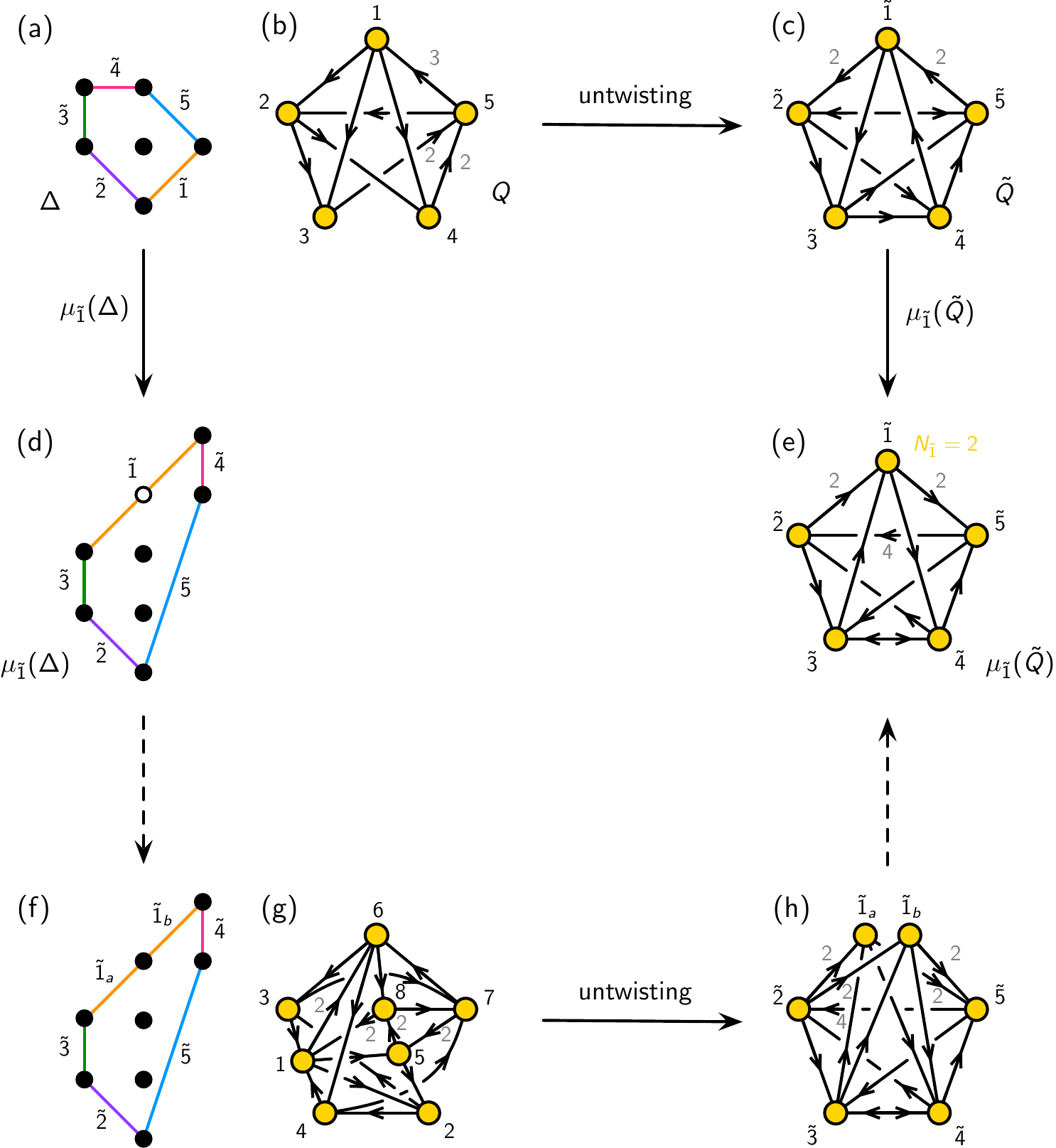}
 }
\caption{Derivation of the twin quiver for the GTP in (d) via two methods, using $dP_{2,(b)}$ as a starting point. The top row uses the algorithm presented in Section \sref{section_quiver_and_polytope_mutations}, while the bottom row uses the one in Section \sref{section_connection_toric_GTP}. Both methods produce the same twin quiver, which is shown in (e).
\label{f_dP2b_mutation2}
}
\end{center}
\end{figure}

We now repeat the analysis, but starting instead from the other toric phase, $dP_{2,(b)}$. All steps are shown in \fref{f_dP2b_mutation2}. On the top row, we observe that untwisting $dP_{2,(b)}$ results in the quiver in $(c)$ which is identical to the one in \fref{f_dP2a_mutation2} (c), with the exception of a bidirectional arrow connecting nodes $\tilde{2}$ and $\tilde{5}$. Mutating this quiver on node $\tilde{1}$ produces the twin quiver for the GTP, which is shown in $(e)$. As in the previous example, the superpotential for this theory is also known. Interestingly, this quiver only differs from the one in \fref{f_dP2a_mutation2} by the presence of a bidirectional arrow between nodes $\tilde{3}$ and $\tilde{4}$. As in the previous example, the alternative method shown in the bottom row produces the same result.

%=================================================================
%=================================================================

We conclude that twin quivers for a given GTP constructed using different toric phases of $Q$ differ by bidirectional arrows. While we reached this conclusion in terms of examples, it can be understood on general grounds as follows. Here we focus on the approach on the bottom row of Figures \ref{f_dP2a_mutation2} and \ref{f_dP2b_mutation2}. After replacing the GTP by a toric diagram, let us refer to the two toric phases associated to it as $Q_{\rm{T},(a)}$ and $Q_{\rm{T},(b)}$. The (sequence of) Seiberg duality transformation(s) connecting $Q_{\rm{T},(a)}$ and $Q_{\rm{T}(b)}$ corresponds to a reorganization of some of the zig-zag paths that preserves their homology \cite{Hanany:2005ss}. When doing so, the intersections between zig-zags change, but they appear/disappear in pairs, with opposite signs. Since nodes in $\tilde{Q}_{\rm{T},(a)}$ and $\tilde{Q}_{\rm{T}(b)}$ correspond to zig-zags, they only differ by bidirectional arrows. This property is preserved when nodes are combined into higher rank ones to form $\tilde{Q}_{\rm{GTP},(a)}$ and $\tilde{Q}_{\rm{GTP}(b)}$.

%=================================================================
\section{The generalized $s$-rule and SUSY breaking in the twin quiver} 
%=================================================================

\label{section_s-rule}

In this section we explain how twin quivers elegantly capture the generalized $s$-rule of GTPs. This observation not only provides a practical tool, but also gives additional support for the correspondence between twin quivers and GTPs. As we will stress, our understanding of this issue is not yet complete. In particularly, currently we do not always know the superpotentials of non-toric twin quivers for general GTPs. As we mentioned earlier, one exception is when \ the non-toric quivers can be obtained via mutations of toric phases. Having said that, we feel that the ideas presented in this section are worth presenting in their current form, since they may inspire further progress.

We propose that violation of the $s$-rule by the GTP corresponds to ordinary SUSY breaking in the twin quiver. The latter corresponds to the presence of a node $\tilde{i}$ in the quiver with $N_{f,\tilde{i}}<N_{c,\tilde{i}}$. If we further mutate the quiver on such a node, its rank becomes negative. Generically, a sequence of quiver mutations might be necessary in order to make the SUSY breaking nature of a twin quiver manifest. In terms of the twin quiver, the $s$-rules becomes:
\begin{itemize}
\item{\bf Rule 1:} a GTP preserves SUSY if there is no duality frame of the corresponding twin quiver that simultaneously has positive and negative ranks.
\end{itemize}
The simultaneous presence of positive and negative ranks corresponds to the coexistence of branes and anti-branes.

%=================================================================
\subsection{The basic building blocks of GTPs} 
%=================================================================

GTPs can be tessellated by two types of elementary building blocks: triangular and trapezium GTPs with only black dots at their corners \cite{Benini:2009gi}. The generalized $s$-rule can then be phrased as conditions that these basic constituents need to satisfy in order for SUSY to be preserved. Below we discuss how the notion of SUSY breaking in the twin quiver captures the $s$-rule for these basic GTPs.

%=================================================================
\subsubsection{Triangles} 
%=================================================================

\begin{figure}[H]
\begin{center}
\resizebox{0.5\hsize}{!}{
  \includegraphics[trim=0mm 0mm 0mm 0mm, width=1.3in]{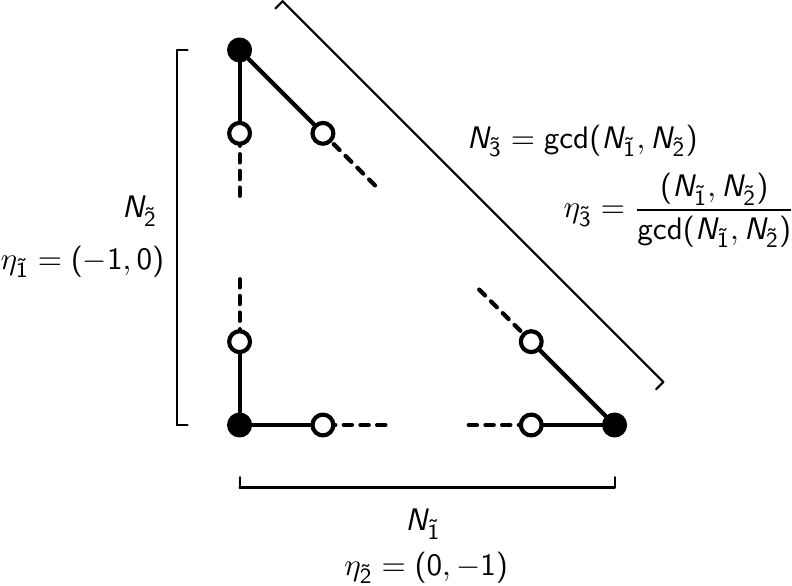}
}
\caption{General triangular GTP.
\label{triangular_GTP}
}
\end{center}
\end{figure}

Let us first consider triangular GTPs. Without loss of generality, we can make sides $\tilde{1}$ and $\tilde{2}$ perpendicular, as shown in \fref{triangular_GTP}.

%=================================================================
%=================================================================

\begin{figure}[H]
\begin{center}
\resizebox{0.55\hsize}{!}{
  \includegraphics[trim=0mm 0mm 0mm 0mm, width=1in]{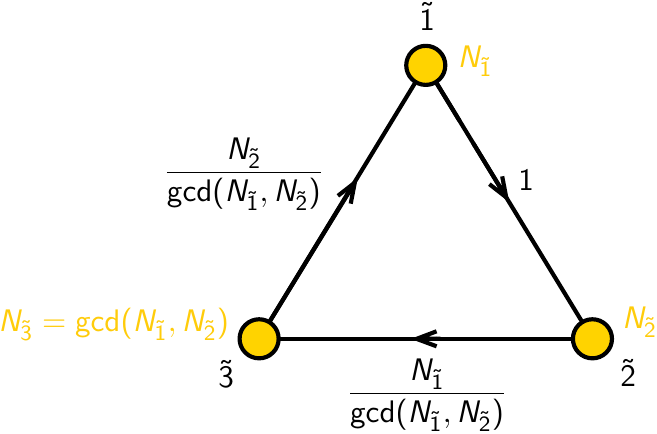}
}
\caption{Twin quiver for a general triangular GTP.
\label{twin_quiver_triangular_GTP}
}
\end{center}
\end{figure}

We therefore have
\beq
\begin{array}{cclcl}
\eta_{\tilde{1}} & = & (-1,0) & \ \ \ \ & N_{\tilde{1}} \\[.15cm]
\eta_{\tilde{2}} & = & (0,-1) & \ \ \ \ & N_{\tilde{2}} \\[.15cm]
\eta_{\tilde{3}} & = & {(N_{\tilde{1}},N_{\tilde{2}})\over \gcd(N_{\tilde{1}},N_{\tilde{2}})} & \ \ \ \ & N_{\tilde{3}}= \gcd(N_{\tilde{1}},N_{\tilde{2}})
\end{array}
\eeq
where $N_{\tilde{3}}$ follows from the fact that side $\tilde{3}$ of the triangle intersects additional lattice points if $N_{\tilde{1}}$ and $N_{\tilde{2}}$ are not coprime. With this information, we use the method in Section \sref{section_direction_construction_quiver_for_GTP} to derive the twin quiver for a general GTP of this form, which we show in \fref{twin_quiver_triangular_GTP}.

%=================================================================
%=================================================================

\begin{figure}[H]
\begin{center}
\resizebox{0.65\hsize}{!}{
  \includegraphics[trim=0mm 0mm 0mm 0mm, width=1in]{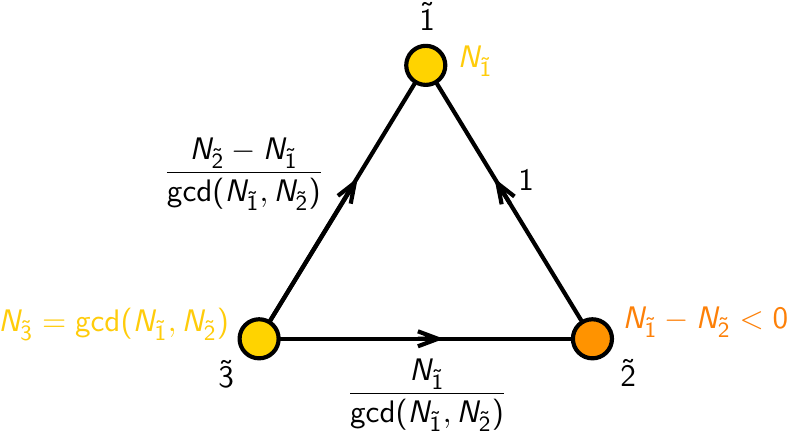}
}
\caption{Mutating the twin quiver in \fref{twin_quiver_triangular_GTP} on node $\tilde{2}$, we obtain a quiver with both positive and negative ranks. 
\label{twin_quiver_triangular_GTP_negative_ranks}
}
\end{center}
\end{figure}

As expected, the quiver is free of anomalies, namely every node has an equal number of incoming and outgoing arrows. Without loss of generality, we can assume that $N_{\tilde{2}} \geq N_{\tilde{1}}$. The number of flavors for node $\tilde{2}$ is equal to $N_{\tilde{1}}$. Therefore, if the inequality is strict, i.e. if $N_{\tilde{2}} > N_{\tilde{1}}$, node $\tilde{2}$ has $N_{f,\tilde{2}}<N_{c,\tilde{2}}$ and the quiver dynamically breaks SUSY. If we formally apply a quiver mutation at node $\tilde{2}$, we obtain the quiver in \fref{twin_quiver_triangular_GTP_negative_ranks}, which simultaneously has positive and negative ranks. Arrows connected to a negative rank node should be interpreted as going in the opposite direction. This phenomenon will also appear in some of the subsequent examples.

%=================================================================
%=================================================================

We conclude that SUSY is only preserved if $N_{\tilde{1}}=N_{\tilde{2}}=N_{\tilde{3}}$. This is precisely the $s$-rule condition for triangular GTPs derived in \cite{Benini:2009gi}.

%=================================================================
\subsubsection{Trapeziums} 
%=================================================================

\label{section_trapeziums}

\begin{figure}[H]
\begin{center}
\resizebox{0.5\hsize}{!}{
  \includegraphics[trim=0mm 0mm 0mm 0mm, width=1.3in]{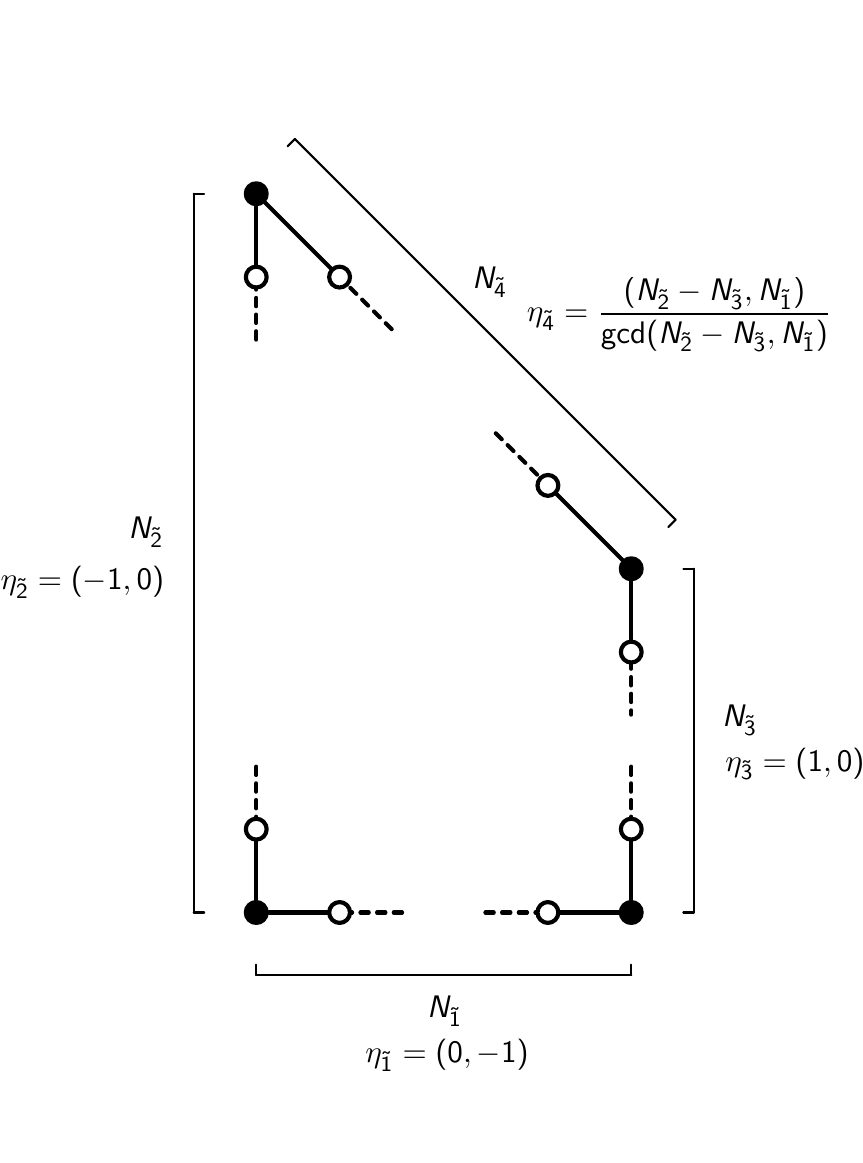}
}
\caption{General trapezium GTP.
\label{trapezium_GTP}
}
\end{center}
\end{figure}

Trapezium GTPs can always be taken to the form in \fref{trapezium_GTP}. The normal vectors and numbers of edges in each of the sides are
\beq
\begin{array}{cclcl}
\eta_{\tilde{1}} & = & (0,-1) & \ \ \ \ & N_{\tilde{1}} \\[.15cm] 
\eta_{\tilde{2}} & = & (-1,0) & \ \ \ \ & N_{\tilde{2}} \\[.15cm]
\eta_{\tilde{3}} & = & (1,0) & \ \ \ \ & N_{\tilde{3}} \\[.15cm]
\eta_{\tilde{4}} & = & {(N_{\tilde{2}}-N_{\tilde{3}},N_{\tilde{1}}) \over  \gcd(N_{\tilde{2}}-N_{\tilde{3}},N_{\tilde{1}}) } & \ \ \ \ & N_{\tilde{4}} = \gcd(N_{\tilde{2}}-N_{\tilde{3}},N_{\tilde{1}})
\end{array}
\eeq

%=================================================================
%=================================================================

\begin{figure}[H]
\begin{center}
\resizebox{0.7\hsize}{!}{
  \includegraphics[trim=0mm 0mm 0mm 0mm, width=1in]{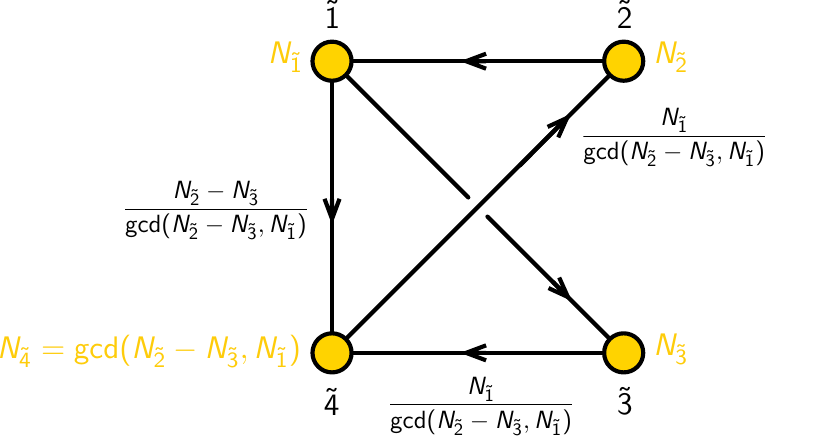}
}
\caption{Twin quiver for a general trapezium GTP.
\label{twin_quiver_trapezium_GTP}
}
\end{center}
\end{figure}

Using the method in Section \sref{section_direction_construction_quiver_for_GTP} we obtain the quiver shown in \fref{twin_quiver_trapezium_GTP}.

%=================================================================
%=================================================================

\begin{figure}[H]
\begin{center}
\resizebox{0.7\hsize}{!}{
  \includegraphics[trim=0mm 0mm 0mm 0mm, width=1in]{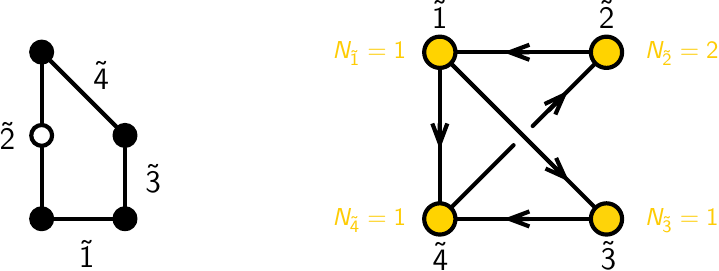}
}
\caption{Trapezium GTP with $N_{\tilde{2}}=2$ and $N_{\tilde{1}}=N_{\tilde{3}}=N_{\tilde{4}}=1$ and the corresponding twin quiver.
\label{twin_quiver_trapezium_example_1}
}
\end{center}
\end{figure}

The $s$-rule for general trapeziums is rather involved, so we will focus on explicit examples. We will see that the treatment of trapeziums leads to a new rule that complements the {\bf Rule 1} introduced above. Let us first consider $N_{\tilde{2}}=2$ and $N_{\tilde{1}}=N_{\tilde{3}}=N_{\tilde{4}}=1$. This example is SUSY and has been considered in \cite{Benini:2009gi}. \fref{twin_quiver_trapezium_example_1} shows the GTP and twin quiver in this case.

%=================================================================
%=================================================================

\begin{figure}[H]
\begin{center}
\resizebox{0.9\hsize}{!}{
  \includegraphics[trim=0mm 0mm 0mm 0mm, width=4in]{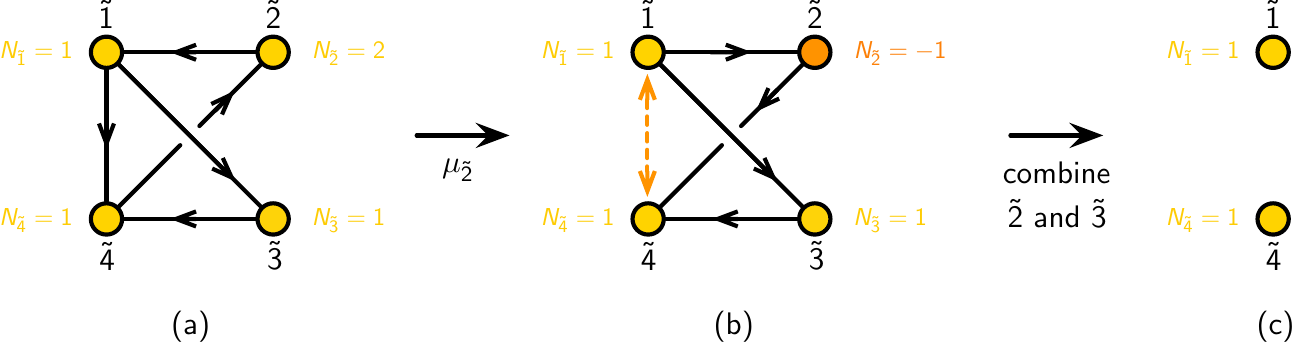}
}
\caption{Mutation and node condensation on the twin quiver in \fref{twin_quiver_trapezium_example_1}.
\label{twin_quiver_trapezium_example_1_mutations}
}
\end{center}
\end{figure}

From the quiver, we immediately see that $N_{f,\tilde{2}} < N_{c,\tilde{2}}$, so this would naively result in SUSY breaking. Let us consider this example more carefully as shown in \fref{twin_quiver_trapezium_example_1_mutations}. $(a)$ shows again the twin quiver for this GTP and $(b)$ shows the result of mutating it on node $\tilde{2}$. When going from $(a)$ to $(b)$, we have deleted the chiral field connecting nodes $\tilde{1}$ and $\tilde{4}$. This assumes that the superpotential of the theory in $(b)$ contains a cubic coupling associated to the $(\tilde{1},\tilde{4},\tilde{2})$ loop in the quiver. If such a term is present, after the mutation it would become a mass term combining the field from $\tilde{1}$ to $\tilde{4}$ and a meson that goes in the opposite direction, and both would disappear from the quiver, as we assumed. If this term was not present, the twin quiver in $(b)$ would contain a bidirectional arrow between nodes $\tilde{1}$ and $\tilde{4}$, which we show as a dashed orange arrow. Which of the two situations is realized in this case affects the final quiver but does not change our conclusion regarding SUSY breaking, so we leave this interesting question for future work. In what follows, we will always assume that such mass terms are present and remove the massive fields. In all the examples considered in the paper, the presence of such fields may modify the final quiver but does not affect our diagnostic of SUSY breaking. However, such fields are important for determining whether SUSY is broken in more involved examples, since omitting them we might incorrectly conclude that $N_{f,\tilde{i}}< N_{c,\tilde{i}}$ for some node of the twin quiver.

As expected, the mutation on node $\tilde{2}$ produces a negative $N_{\tilde{2}}$ which, combined with the other positive ranks, would seem to indicate SUSY breaking. However, we know that this particular GTP is SUSY preserving. This example suggests that we have to add a new rule:

\begin{itemize}
\item{\bf Rule 2:} Combine nodes with vanishing intersection and ranks of opposite signs.
\end{itemize}

\noindent We will say that a GTP breaks SUSY if there is a duality frame in which ranks of opposite signs coexist even after applying {\bf Rule 2}.

When combining nodes, we simply add their ranks. Moreover, since their mutual intersection is zero, the intersections with the rest of the nodes in the quiver differ at most by a sign. In this example, these nodes are $\tilde{2}$ and $\tilde{3}$. They have opposite ranks and hence cancel each other. This operation is shown in \fref{twin_quiver_trapezium_example_1_mutations} $(c)$. The final result only has ranks of a given sign, in this case positive, so we conclude that the GTP is indeed SUSY preserving. As mentioned earlier, whether we end up with two isolated nodes as in the figure or with a bidirectional arrow connecting them depends on detailed structure of the superpotential. In Section \sref{section_further_thoughts_brane_motion} we will comment on the brane interpretation of the new rule. 

%=================================================================
%=================================================================

\begin{figure}[H]
\begin{center}
\resizebox{0.7\hsize}{!}{
  \includegraphics[trim=0mm 0mm 0mm 0mm, width=1in]{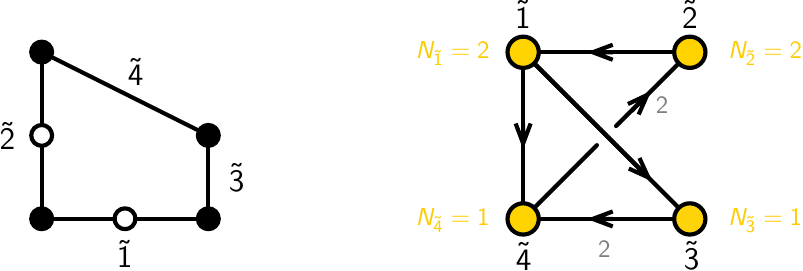}
}
\caption{Trapezium GTP with $N_{\tilde{1}}=2$, $N_{\tilde{2}}=2$, $N_{\tilde{3}}=1$ and $N_{\tilde{4}}=1$ (we only show its boundary) and the corresponding twin quiver.
\label{twin_quiver_trapezium_example_2}
}
\end{center}
\end{figure}

Let us now consider a SUSY breaking trapezium, which was also discussed in \cite{Benini:2009gi}. In this case, $N_{\tilde{1}}=2$, $N_{\tilde{2}}=2$, $N_{\tilde{3}}=1$ and $N_{\tilde{4}}=1$. \fref{twin_quiver_trapezium_example_2} shows the GTP and twin quiver for this model, which are specializations of the general ones in Figures \ref{trapezium_GTP} and \ref{twin_quiver_trapezium_GTP}.

%=================================================================
%=================================================================

\begin{figure}[H]
\begin{center}
\resizebox{1\hsize}{!}{
  \includegraphics[trim=0mm 0mm 0mm 0mm, width=4in]{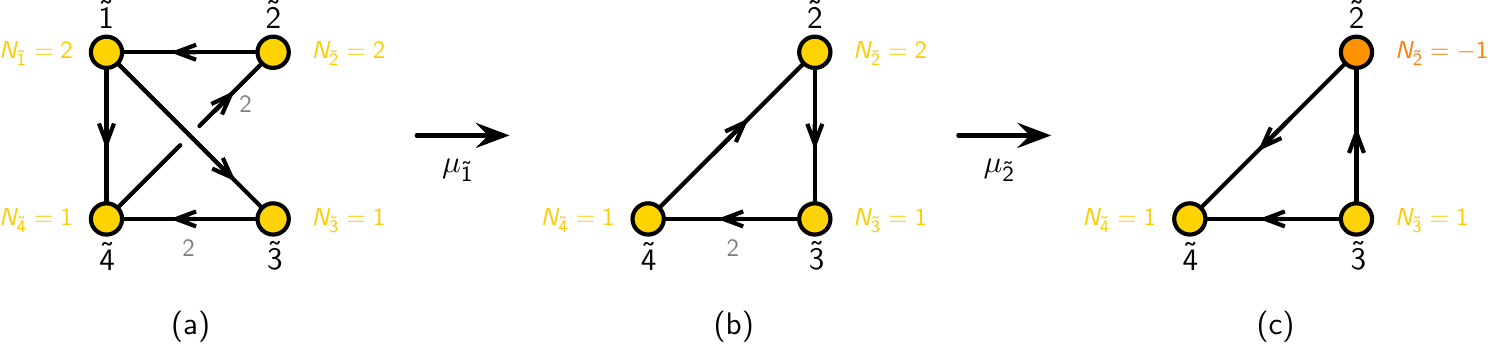}
}
\caption{Sequence of mutations on the twin quiver in \fref{twin_quiver_trapezium_example_2}.
\label{twin_quiver_trapezium_example_2_mutations}
}
\end{center}
\end{figure}

We now consider the sequence of mutations shown in \fref{twin_quiver_trapezium_example_2_mutations}. At the starting point, node $\tilde{1}$ has $N_{f,\tilde{1}}= N_{c,\tilde{1}}$, so it disappears upon the mutation that takes from $(a)$ to $(b)$. The next mutation on node $\tilde{2}$ produces a negative rank that cannot be eliminated by {\bf Rule 2}. We therefore conclude that the GTP is SUSY breaking.

%=================================================================
%=================================================================

%=================================================================
\subsubsection{A more general example} 
%=================================================================

\begin{figure}[H]
\begin{center}
\resizebox{0.7\hsize}{!}{
  \includegraphics[trim=0mm 0mm 0mm 0mm, width=1in]{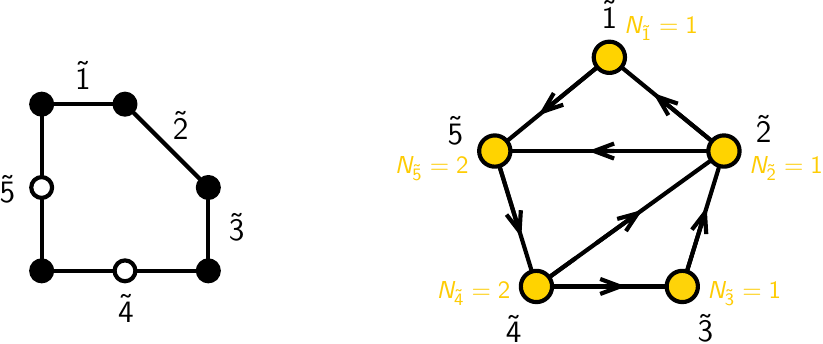}
}
\caption{A general GTP (we only show its boundary) and the corresponding twin quiver.
\label{twin_quiver_general_GTP_example}
}
\end{center}
\end{figure}

Let us finally consider an example that is not an elementary triangle or trapezium. \fref{twin_quiver_general_GTP_example} shows the GTP and corresponding twin quiver. This GTP was considered in \cite{Benini:2009gi}.

%=================================================================
%=================================================================

\begin{figure}[H]
\begin{center}
\resizebox{.7\hsize}{!}{
  \includegraphics[trim=0mm 0mm 0mm 0mm, width=4in]{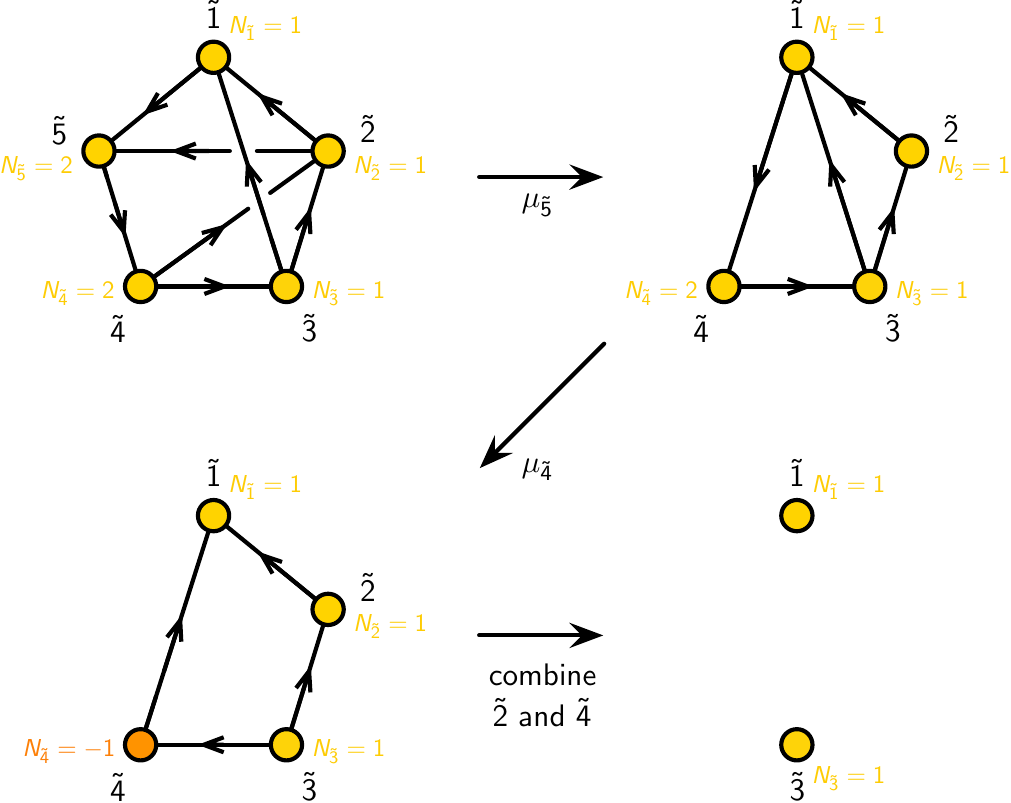}
}
\caption{Sequence of mutations and node condensation on the twin quiver in \fref{twin_quiver_general_GTP_example}.
\label{twin_quiver_general_GTP_example_mutations}
}
\end{center}
\end{figure}

\fref{twin_quiver_general_GTP_example_mutations} shows a sequence of mutations and node condensation terminating in a SUSY configuration. While we only present one possible manipulation of the twin quiver, it is easy to convince ourselves that similar sequences of transformations making SUSY breaking explicit do not exit. We therefore conclude that this GTP is SUSY, in agreement with \cite{Benini:2009gi}. 

%=================================================================
%=================================================================

The results for this model are encouraging. Having said that, it is natural to expect that understanding the generalized $s$-rule in terms of twin quivers for general GTPs, beyond the basic triangles and trapeziums, may require knowing not only their superpotential, but also how tessellations of the GTP are captured by the quivers.

%=================================================================
\subsection{Further thoughts on brane motion and polytope mutation} 
%=================================================================

\label{section_further_thoughts_brane_motion}

The connection between polytope mutation and transformations of brane setups was discussed in Section \sref{section_polytope_mutation_and_7-branes}. This correspondence underlies the twin quiver perspective on the $s$-rule. Having said that, the way the brane transformations were presented in \cite{Benini:2009gi} might appear superficially different from ours. In this section we briefly go over the analysis presented in \cite{Benini:2009gi} for a couple of examples and explain how their transformations are indeed equivalent to our approach. While the connection might be clear to many readers, it is instructive to discuss it for completeness.

\begin{figure}[H]
\begin{center}
\resizebox{0.9\hsize}{!}{
  \includegraphics[trim=0mm 0mm 0mm 0mm, width=1.5in]{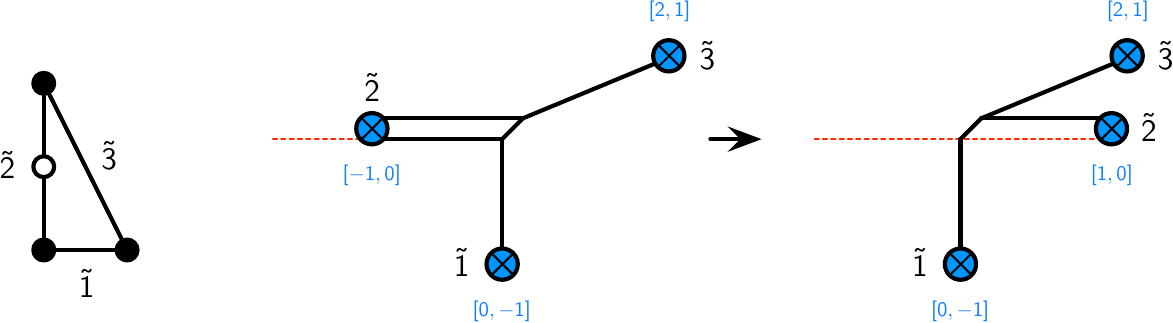}
}
\caption{A GTP that violates the $s$-rule and the corresponding brane configuration. SUSY breaking becomes manifest in the final figure.
\label{fwebs01}
}
\end{center}
\end{figure}

\fref{fwebs01} shows one of the simplest SUSY breaking triangular GTPs. Next to it, we show the corresponding brane configuration. Two D5-branes stretch from the $[-1,0]$ 7-brane to the web. Moving this 7-brane to the right, we are left with an anti D5-brane due to the brane creation/annihilation mechanism. The full configuration thus breaks SUSY.

%=================================================================
%=================================================================

In the final configuration, notice that a $(1,1)$ 5-branes becomes a vertical NS5-brane when crossing the branch cut. This is the main difference between the approach in \cite{Benini:2009gi} and the one that we used, which we described in Section \sref{section_polytope_mutation_and_7-branes}. In \cite{Benini:2009gi}, when a 5-brane crosses a branch cut, it is transformed by the monodromy and changes its slope. This is a schematic way to represent the bending of the 5-brane on the curved background generated by the 7-brane. Instead, we phrased the same process as a change in the charge of the 7-branes when crossing branch cuts. The two viewpoints are equivalent. 

Notice that in this perspective the fact that only a subset of the 7-branes (or equivalently of the 5-brane legs) with either positive or negative $\langle \eta_{\tilde{j}}, \eta_{\tilde{i}}\rangle$ crosses and therefore is affected by the branch cut of the 7-brane $\tilde{j}$ becomes transparent. The intersection number $\langle \eta_{\tilde{j}}, \eta_{\tilde{i}}\rangle$ defined in \eref{intersection_sides} is simply the cross product of the vectors $\eta_{\tilde{j}}$ and $\eta_{\tilde{i}}$, which live on the $xy$-plane. Therefore, the sign of the intersection distinguishes the two half planes separated by the branch cut of $\tilde{j}$. Depending on where we locate the branch cut, only one of the two sets of 5-branes indeed crosses it.

\begin{figure}[H]
\begin{center}
\resizebox{0.9\hsize}{!}{
  \includegraphics[trim=0mm 0mm 0mm 0mm, width=1.5in]{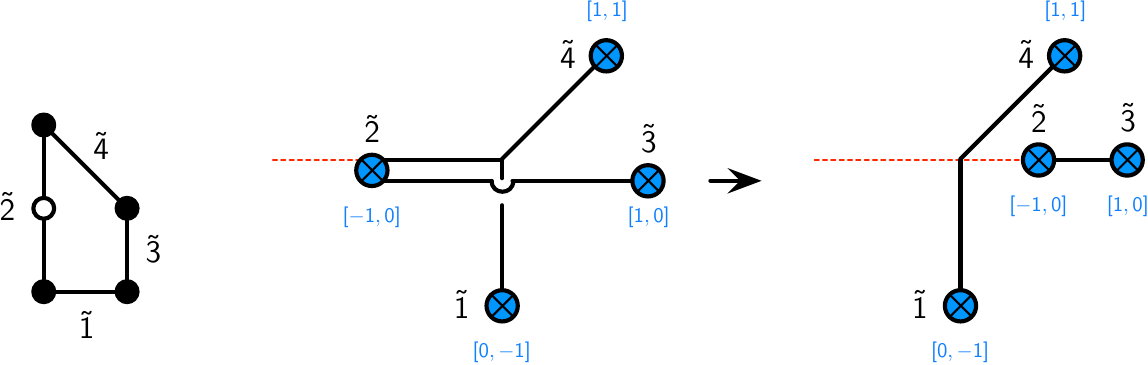}
}
\caption{A GTP that satisfies the $s$-rule and the corresponding brane configuration.
\label{fwebs02}
}
\end{center}
\end{figure}

Let us now consider the model in \fref{fwebs02}. Its brane analysis is very similar to the previous example. In Section \sref{section_trapeziums}, we analyzed the $s$-rule for this model in terms of the twin quiver. Here we would simple like to point out that {\bf Rule 2}, which regards the cancellation of non-intersecting nodes in the quiver, corresponds in this case to branes $\tilde{2}$ and $\tilde{3}$, which decouple from the rest of the setup and can annihilate each other.

%=================================================================
%=================================================================

%=================================================================
\section{Conclusions and future directions} 
%=================================================================

\label{section_conclusions}

We have proposed a democratic treatment of two sets of quiver theories, polytopes and geometries that normally appear in the study of BFTs. We referred to them as the original and twin theories. The two sides of this correspondence are, in the simplest cases, connected by the operation known as untwisting. Only a subset of these objects has been normally considered in the literature. The unified perspective that we advocate gives rise to natural new questions regarding connections between these objects, some of which we explored in this paper. 

With this motivation, we established a correspondence between the mutations of the original polytope and the twin quiver (equivalently, the mutations of the twin polytope and the original quiver). Mutations that result in non-toric phases of the twin quivers are particularly interesting. We proposed that non-toric twin quivers are naturally associated to GTPs and investigated various aspects of such correspondence. The evidence supporting the proposal includes global symmetries, the ability of twin quivers to encode the $s$-rule, and the connection between polytope mutations and transformations of configurations of $(p,q)$-webs suspended from 7-branes. We introduced three different algorithms for deriving twin quivers associated to GTPs. The first one is based on mutations, the second one uses the interpretation of a GTP as a toric diagram as an intermediate step, and a final one computes the twin quiver from basic information about the GTP. The last method has certain limitations, which we discussed. We also studied the relation between twin quivers for GTPs obtained from different toric phases. Our analysis indicates that twin quivers provide a powerful new perspective on GTPs.

There are several directions worth pursuing to confirm the proposed correspondence between non-toric twin quivers and general GTPs and, if correct, to determine what else can be learned from them. Here we mention some of the most interesting ones, which we plan to revisit in the future:

\begin{itemize}
\item We associated twin quivers to full GTPs. It would be interesting to investigate how tessellations of GTPs translate into the language of twin quivers.

\item The current proposal is summarized in \fref{open_question}, where the question mark indicates an unknown entry in the correspondence, the original theory $Q$ for a generic GTP, i.e. one that is not a toric diagram.  Since in the case of ordinary toric diagrams the quiver theory/brane tiling $Q$ captures the BPS quiver of the associated $5d$ theory, we expect that such a theory would also encode the BPS spectrum of the $5d$ theories for generic GTPs. We anticipate that answering this question will require a description of a non-toric $\tilde{Q}$ in terms of a generalization of BFTs/brane tilings to non-toric phases obtained from BFTs by mutations and an extension of untwisting to such constructions. Both problems are extremely interesting in their own right. 

%=================================================================
\begin{figure}[ht!]
\begin{center}
\resizebox{.95\hsize}{!}{
  \includegraphics[trim=0mm 0mm 0mm 0mm, width=1.5in]{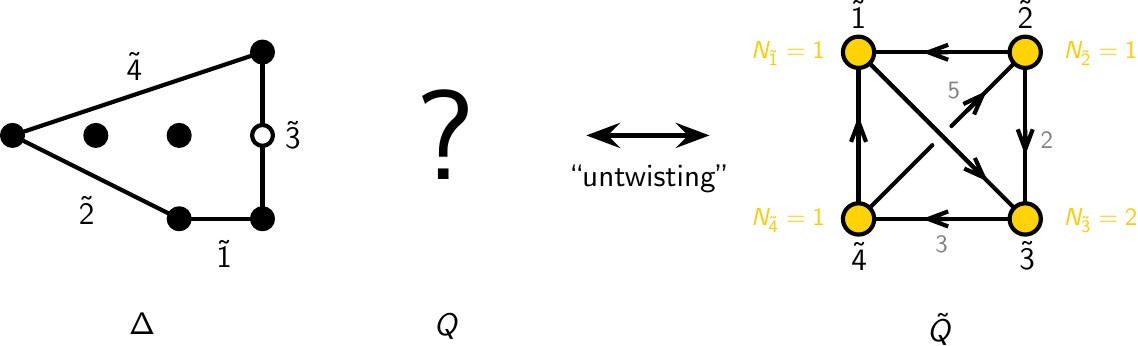}
}
\caption{We have proposed that a general GTP $\Delta$ is associated to a non-toric twin quiver $\tilde{Q}$. What would the corresponding original theory $Q$ be in this case remains an open question. Addressing it would probably require a generalization of untwisting.
\label{open_question}
}
\end{center}
\end{figure}
%=================================================================

\item Untwisting plays a central role in the connection between brane tilings and the mirror configuration of intersecting D6-branes \cite{Feng:2005gw}. Recently, \cite{Bourget:2023wlb} introduced a geometric realization of brane configurations including 7-branes, namely those associated to GTPs, in terms of D6-branes in a Type IIA frame. It would be interesting to explore the connection between this approach and our work, and whether the ideas in \cite{Bourget:2023wlb} can shed light on the question raised in the previous bullet point.

\end{itemize}

While we have primarily focused on its implications for GTPs, we consider that the unified perspective put forward in Section \sref{section_democratic_perspective} and elaborated throughout the paper can have interesting applications in a much wider range of problems.

%======================================================================  
\acknowledgments
%======================================================================  

S.F. would like to thank Diego Rodr\'iguez-G\'omez for stimulating discussions and encouragement. S.F was supported by the U.S. National Science Foundation grants PHY-2112729 and DMS-1854179. 
R.K.-S. is supported by a Basic Research Grant of the National Research Foundation of Korea (NRF-2022R1F1A1073128).
He is also supported by a Start-up Research Grant for new faculty at UNIST (1.210139.01), a UNIST AI Incubator Grant (1.230038.01) and a UNIST Fundamental Science Research Grant (1.220123.01), as well as an Industry Research Project (2.220916.01) funded by Samsung SDS in Korea.  
He is also partly supported by the BK21 Program (``Next Generation Education Program for Mathematical Sciences'', 4299990414089) funded by the Ministry of Education in Korea and the National Research Foundation of Korea (NRF).

%%%%%%%%%%%%%%%%%%%%%%%%%%%%%%%%%%%%%%%%%%%%%%%

\appendix

\bigskip
\bigskip

%=================================================================
\section{Additional details on some of the models}
%=================================================================

\label{appendix_details_models}

In this Appendix we collect additional information on the models discussed in Section \sref{section_twin_quivers_for_different_toric_phases}. We will present the bipartite graphs for the toric phases in Figures \ref{f_dP2a_mutation2} and \ref{f_dP2b_mutation2}. We include the zig-zag paths in order to understand how untwisting connects the different theories. From the bipartite graphs, one can not only recover the quivers that were given in Section \sref{section_twin_quivers_for_different_toric_phases}, but also their superpotentials.

\newpage

%=================================================================
\subsection*{Models on the top row of \fref{f_dP2a_mutation2}}
%=================================================================

%=================================================================
\begin{figure}[H]
\begin{center}
\resizebox{0.9\hsize}{!}{
  \includegraphics[trim=0mm 0mm 0mm 0mm, width=1.5in]{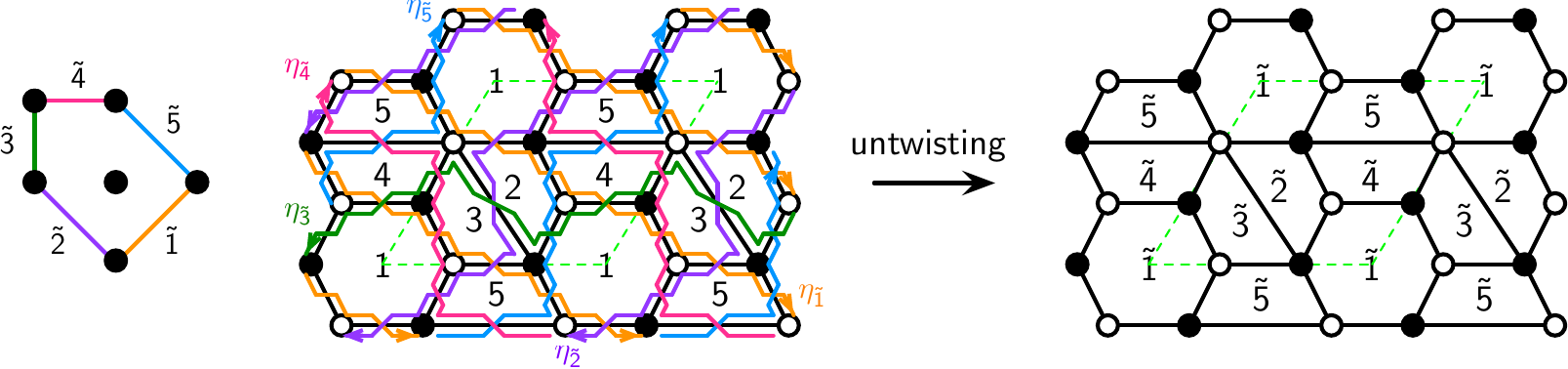}
}
\caption{BFTs for the models on the top row of \fref{f_dP2a_mutation2}.
\label{f28}
}
\end{center}
\end{figure}
%=================================================================

%=================================================================
\subsection*{Models on the bottom row of \fref{f_dP2a_mutation2}}
%=================================================================

%=================================================================
\begin{figure}[H]
\begin{center}
\resizebox{0.9\hsize}{!}{
  \includegraphics[trim=0mm 0mm 0mm 0mm, width=1.5in]{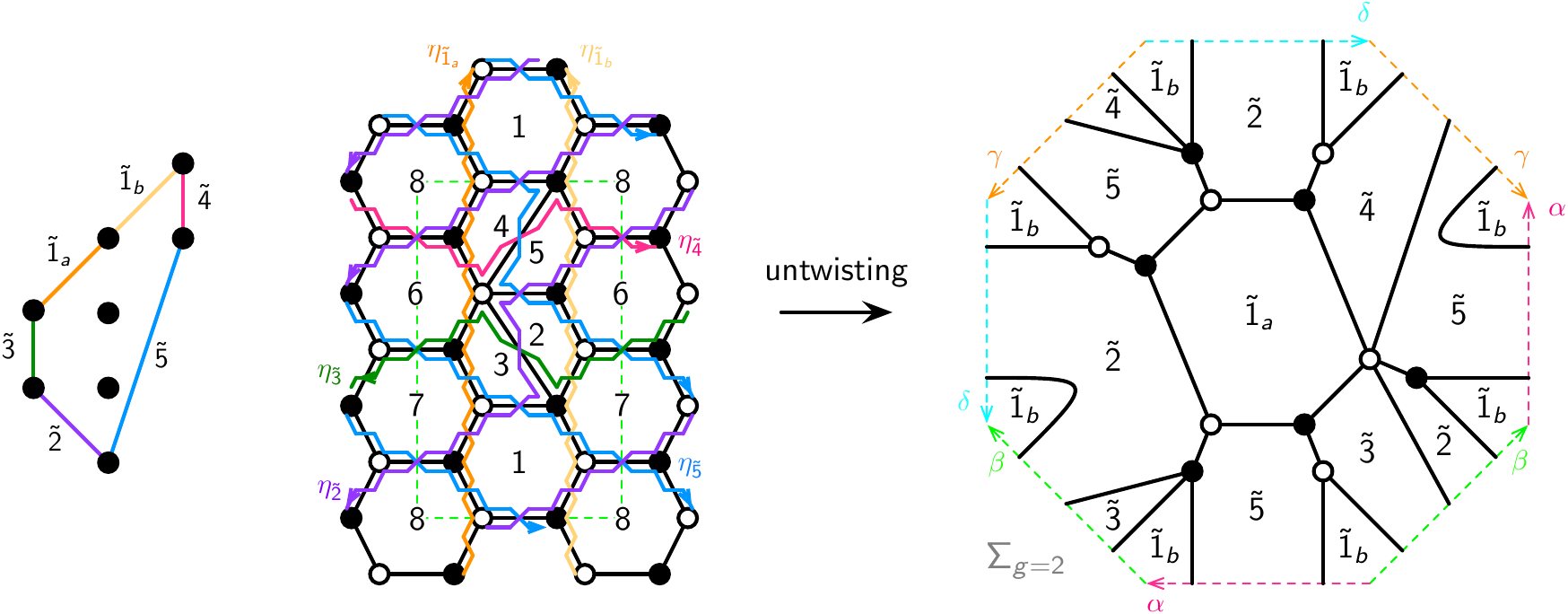}
}
\caption{BFTs for the models on the bottom row of \fref{f_dP2a_mutation2}.
\label{f29}
}
\end{center}
\end{figure}
%=================================================================

%=================================================================
\subsection*{Models on the top row of \fref{f_dP2b_mutation2}}
%=================================================================

%=================================================================
\begin{figure}[H]
\begin{center}
\resizebox{0.9\hsize}{!}{
  \includegraphics[trim=0mm 0mm 0mm 0mm, width=1.5in]{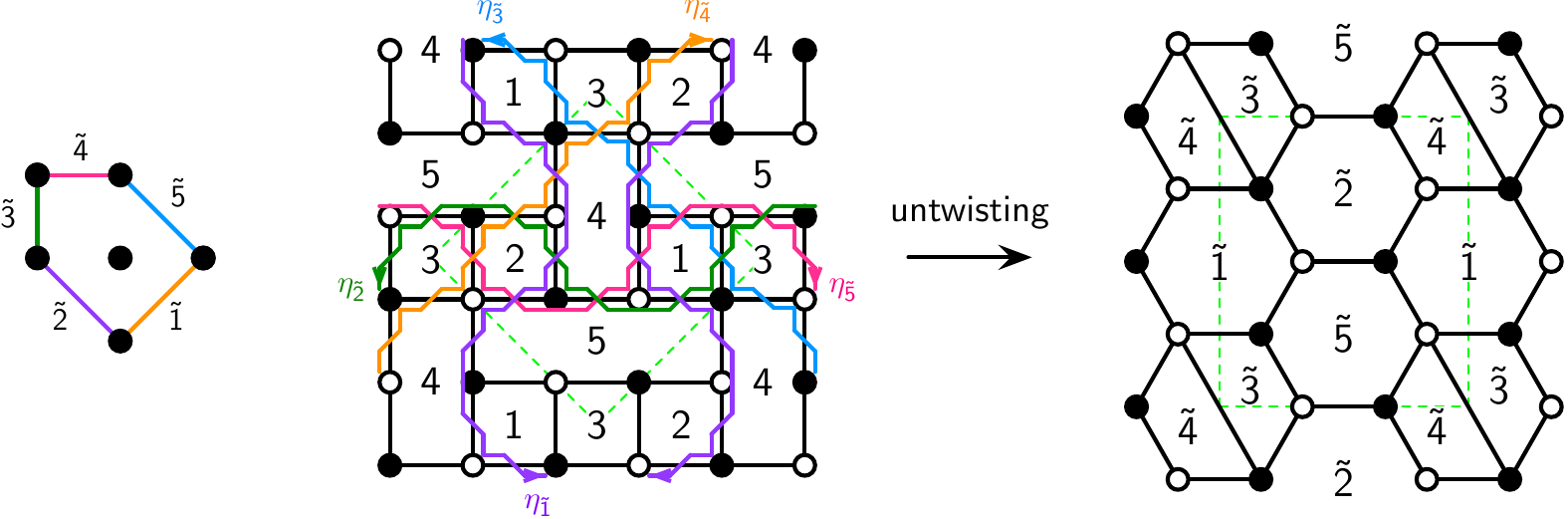}
}
\caption{BFTs for the models on the top row of \fref{f_dP2b_mutation2}.
\label{f30}
}
\end{center}
\end{figure}
%=================================================================

%=================================================================
\subsection*{Models on the bottom row of \fref{f_dP2b_mutation2}}
%=================================================================

%=================================================================
\begin{figure}[H]
\begin{center}
\resizebox{1\hsize}{!}{
  \includegraphics[trim=0mm 0mm 0mm 0mm, width=1.5in]{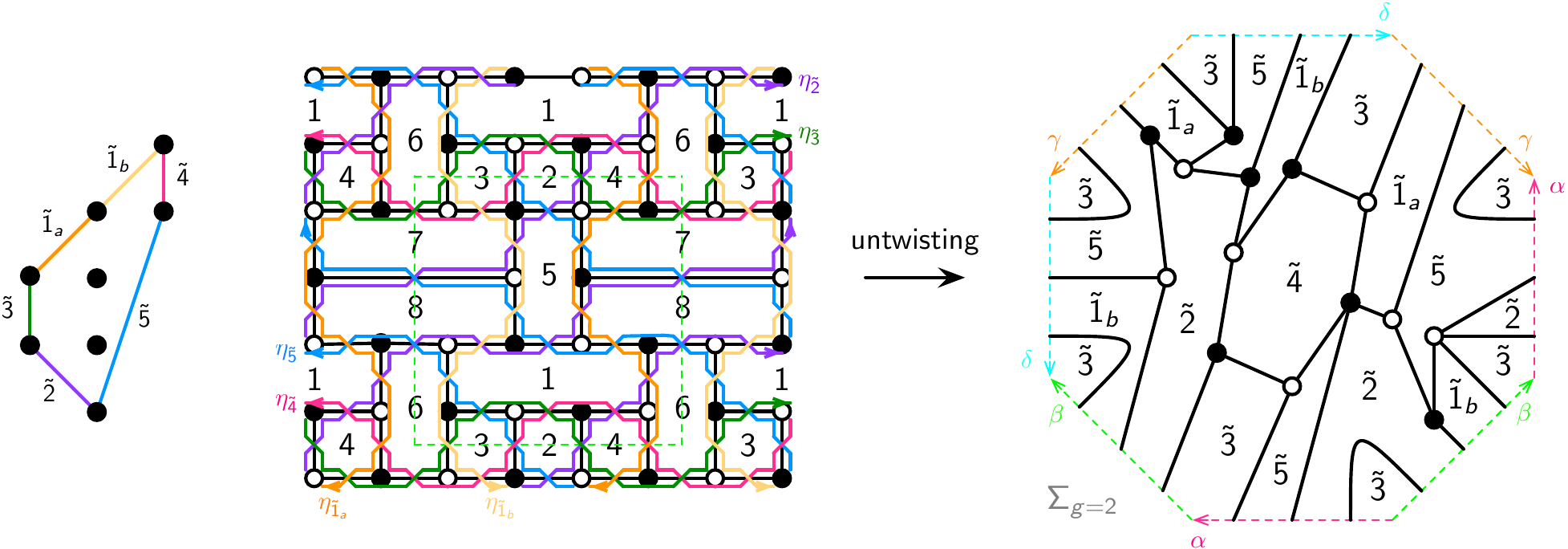}
}
\caption{BFTs for the models on the bottom row of \fref{f_dP2b_mutation2}.
\label{f31}
}
\end{center}
\end{figure}
%=================================================================

%======================================================================
\bibliographystyle{JHEP}
\bibliography{mybib}
%======================================================================

\end{document}

%% file: pref.tex
\newcommand{\be}{\begin{equation}}
\newcommand{\ee}{\end{equation}}
\newcommand{\beq}{\begin{equation}}
\newcommand{\beql}[1]{\begin{equation}\label{#1}}
\newcommand{\eeq}{\end{equation}}
\newcommand{\ba}{\begin{array}}
\newcommand{\ea}{\end{array}}
\newcommand{\bea}{\begin{eqnarray}}
\newcommand{\beal}[1]{\begin{eqnarray}\label{#1}}
\newcommand{\eea}{\end{eqnarray}}
\newcommand{\ben}{\begin{enumerate}}
\newcommand{\een}{\end{enumerate}}
\newcommand{\bean}{\begin{eqnarray*}}
\newcommand{\eean}{\end{eqnarray*}}
\newcommand{\eref}[1]{(\ref{#1})}
\newcommand{\sref}[1]{\S\ref{#1}}

\newcommand{\fref}[1]{Figure \ref{#1}}
\newcommand{\btab}[1]{\begin{tabular}{#1}}
\newcommand{\etab}{\end{tabular}}

\def \Black {\pdfsetcolor{0 0 0 1}}

\newcommand{\comment}[1]{}

\newcommand{\qed}{\nobreak \ifvmode \relax \else
      \ifdim\lastskip<1.5em \hskip-\lastskip
      \hskip1.5em plus0em minus0.5em \fi \nobreak
      \vrule height0.75em width0.5em depth0.25em\fi}